\documentclass[ reprint, superscriptaddress, amsmath,amssymb,aps]{revtex4-2}
\usepackage{graphicx}
\usepackage{dcolumn}
\usepackage{bm}
\usepackage[utf8]{inputenc}
\usepackage{color}
\usepackage{soul}
\begin{document}

\preprint{APS/123-QED}

\title{Long-range magnetic order induced surface state in GdBi and DyBi}

\author{Yevhen Kushnirenko}
\email[]{kushes16@gmail.com}
\affiliation{Division of Materials Science and Engineering, Ames Laboratory, Ames, Iowa 50011, USA}
\affiliation{Department of Physics and Astronomy, Iowa State University, Ames, Iowa 50011, USA}

\author{Lin-Lin Wang}
\affiliation{Division of Materials Science and Engineering, Ames Laboratory, Ames, Iowa 50011, USA}

\author{Zhuoqi Li}
\affiliation{Division of Materials Science and Engineering, Ames Laboratory, Ames, Iowa 50011, USA}
\affiliation{Department of Physics and Astronomy, Iowa State University, Ames, Iowa 50011, USA}

\author{Brinda Kuthanazhi}
\affiliation{Division of Materials Science and Engineering, Ames Laboratory, Ames, Iowa 50011, USA}
\affiliation{Department of Physics and Astronomy, Iowa State University, Ames, Iowa 50011, USA}

\author{Benjamin Schrunk}
\affiliation{Division of Materials Science and Engineering, Ames Laboratory, Ames, Iowa 50011, USA}

\author{Evan O'Leary}
\affiliation{Division of Materials Science and Engineering, Ames Laboratory, Ames, Iowa 50011, USA}
\affiliation{Department of Physics and Astronomy, Iowa State University, Ames, Iowa 50011, USA}

\author{Andrew Eaton}
\affiliation{Division of Materials Science and Engineering, Ames Laboratory, Ames, Iowa 50011, USA}
\affiliation{Department of Physics and Astronomy, Iowa State University, Ames, Iowa 50011, USA}

\author{P. C. Canfield}
\affiliation{Division of Materials Science and Engineering, Ames Laboratory, Ames, Iowa 50011, USA}
\affiliation{Department of Physics and Astronomy, Iowa State University, Ames, Iowa 50011, USA}

\author{Adam Kaminski}
\email[]{kaminski@ameslab.gov}
\affiliation{Division of Materials Science and Engineering, Ames Laboratory, Ames, Iowa 50011, USA}
\affiliation{Department of Physics and Astronomy, Iowa State University, Ames, Iowa 50011, USA}

\begin{abstract}
    The recent discovery of unconventional surface-state pairs, which give rise to Fermi arcs and spin textures, in antiferromagnetically ordered rare-earth monopnictides attracted the interest in these materials. We use angle-resolved photoemission spectroscopy (ARPES) measurements in conjunction with density functional theory (DFT) calculations to investigate the evolution of the electronic structure of GdBi and DyBi. We find that new surface states, including a Dirac cone, emerge in the AFM state. However, they are located along a direction in momentum space that is different than what was found in NdSb, NdBi, and CeBi \cite{kushnirenko2022rare,SchrunkNature2022,honma2023antiferromagnetic,kushnirenko2023directional,honma2023unusual}. The observed changes in the electronic structure are consistent with the presence of AFM-II-A type order.
\end{abstract}

\maketitle


\section{Introduction}
The rare-earth monopnictides family of materials crystallizes in a simple NaCl structure. Many members of this family order antiferromagnetically (AFM) \cite{Tsuchida_1965,nereson_1971,bartholin1979hydrostatic,KumigashiraPRB1996}, and some of them have complex phase diagrams with multiple AFM phases \cite{rossat1983magnetic,wiener2000magnetic,Kuroda_2020,kuthanazhi2022magnetisation,kushnirenko2024}. Recent predictions \cite{GuoNPJ2017,li2017predicted,DuanCommPhys2018,ZhuPRB2020,zeng2015topological,wang2022multi} of the existence of topological states that result from the inversion between 6p band of pnictogen element and 5d band of rare-earth element in some of these materials inspired more experimental studies, including angle-resolved photoemission spectroscopy (ARPES). These studies demonstrated not only topological Dirac states \cite{zeng2016compensated,Niu2016Presence,lou2017evidence,Kuroda_2018,li2018tunable,SatoCeBi,sakhya2022behavior,SakhyaNdSb2022} but also the development of interesting surface states (SS) in NdBi \cite{kushnirenko2022rare,SchrunkNature2022,honma2023antiferromagnetic}, NdSb \cite{kushnirenko2022rare,kushnirenko2023directional,honma2023unusual}, and CeBi \cite{kushnirenko2022rare, kushnirenko2024} upon AFM transition. These states undergo Kaminski-Canfield (KC) band splitting that leads to the formation of spin textured Fermi arcs.

GdBi and DiBi were also shown to have an AFM phase. The Neel temperature is $T_N$~=~27.5~K for GdBi \cite{nereson_1971,li1996electrical,dwari2023large} and $T_N$~=~11~K for DyBi \cite{Tsuchida_1965,hulliger1980magnetic,zhao2022multiple}. The AFM structure of these materials is different from that of NdBi, NdSb, and CeBi. Neutron diffraction studies \cite{nereson_1971,nereson1972neutron} have shown, that in both GdBi and DyBi the magnetic moments are aligned in type-II AFM structure. However, neutron diffraction measurements cannot distinguish AFM-II-A (Fig.~2(b)) from AFM-II-B (Fig.~2(c)) order \cite{li1955magnetic}. Similarly to several other rare-earth monopnictides, the DFT calculations for GdBi in the paramagnetic (PM) phase \cite{DuanCommPhys2018,li2018tunable,dwari2023large} predict a band inversion between Bi 6p and Gd 5d bands along $\Gamma-X$ direction. This band inversion should lead to the formation of one SS Dirac cone at the $\overline{\Gamma}$-point and two cones at $\overline{M}$-point of the 2D Brillouin zone (BZ) \cite{Kuroda_2018,honma2023antiferromagnetic}. An earlier synchrotron-based ARPES measurements \cite{li2018tunable} demonstrated the presence of these Dirac states.

In this work, we investigate the evolution of GdBi and DyBi electronic structures upon cooling through their AFM transitions using high-resolution laser-based ARPES. We demonstrate the development of Dirac states in the magnetically ordered phase near the $\overline{X}$-point.

\section{Experimental details}
Single crystals of GdBi and DyBi were grown out of In flux \cite{Canfield1992Growth}. The elements with an initial stoichiometry of R$_4$Bi$_4$In$_{96}$ (R = Gd, Dy) were put into a fritted alumina crucible \cite{Canfield2016Use} (sold as Canfield crucible set \cite{FrittURL}) and sealed in a fused silica tube under partial pressure of argon. The prepared ampoules were heated to 1150$^\circ$~C over 5 hours and held there for 5 hours. This was followed by a slow cooling to 700$^\circ$~C (for GdBi) and 400$^\circ$~C (for DyBi) over 100 hours and decanting of the excess flux through the frit using a centrifuge.\cite{Canfield_2019} The cubic crystals obtained were stored and handled in a glovebox under nitrogen atmosphere.

Most of the ARPES data were collected using vacuum ultraviolet (VUV) laser ARPES spectrometer that consists of a Scienta DA30 electron analyzer, tunable picosecond Ti:Sapphire oscillator paired with fourth-harmonic generator \cite{jiang2014tunable} and 3.5~eV continuous laser paired with second-harmonic generator. Data were collected with 6.2-7~eV photon energy. Angular resolution was set at $\sim$ 0.1$^{\circ}$ and 1$^{\circ}$, along and perpendicular to the direction of the analyzer slit, respectively, and the energy resolution was set at 2~meV. The VUV laser beam was set to vertical polarization. The diameter of the photon beam on the sample was $\sim 15\,\mu$m. The measurements at photon energy of 21.2~eV were carried out using R8000 analyzer and GammaData helium discharge lamp with custom focusing optics. The diameter of the photon beam on the sample was $\sim1 $mm. Samples were cleaved \textit{in-situ} along (001) plane, usually producing very flat, mirror-like surfaces. The measurements were performed at a base pressure lower than 2$\times$10$^{-11}$ Torr.

\begin{figure*}[t]
    \includegraphics[width=7 in]{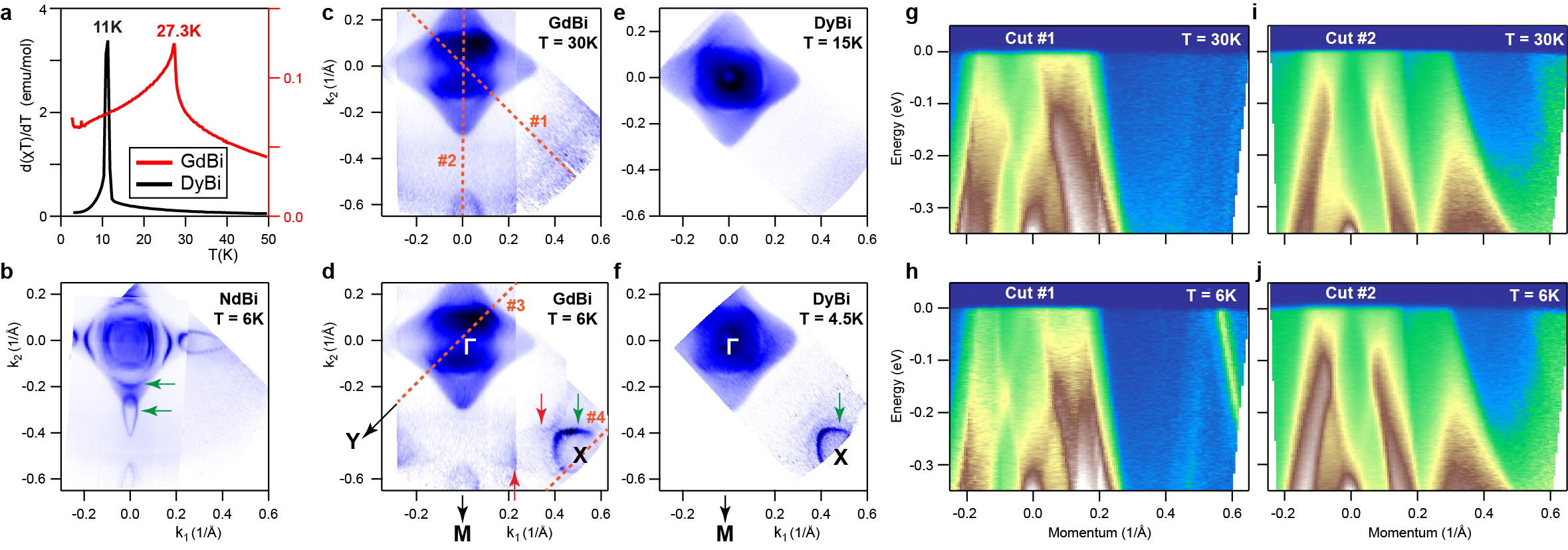}
    \caption{(a) Results of magnetization measurements ($d(\chi T)/dT$) of GdBi and DyBi.
    (b) Fermi surface maps of NdBi measured in the AFM state.
    (c) and (d) Fermi surface maps of GdBi measured in the PM and AFM phases, respectively. 
    (e) and (f) Fermi surface maps of DyBi measured in the PM and AFM phases, respectively. 
    (g) and (h) Band dispersion measured in GdBi in the PM and AFM phases, respectively, along $\overline{\Gamma}-\overline{X}$ direction (marked as \#1 in (c)).
    (i) and (j) Band dispersion measured in GdBi in the PM and AFM phases, respectively, along $\overline{\Gamma}-\overline{M}$ direction (marked as \#2 in (c)).
    The cuts along \#3 and \#4 are shown in Fig.~3.
    The green arrows in (b, d, f) mark SS states that appear in the AFM phase. The red arrows in (d) mark the star-shaped bulk pocket, which appears in the AFM phase.}
\end{figure*}

\section{Results and Discussion}

Temperature-dependent magnetic susceptibility measurements made in H~=~1~kOe were made on GdBi and NdBi. The $d(\chi T)/dT$ data (fig. 1a) show that there is clear AFM ordering below 27.3~K and 11~K for GdBi and DyBi, respectively. These results are in agreement with earlier reported $T_N$ values.
\cite{nereson_1971, li1996electrical, dwari2023large, Tsuchida_1965, hulliger1980magnetic, zhao2022multiple}

\begin{figure}[b]
    \includegraphics[width=1\linewidth]{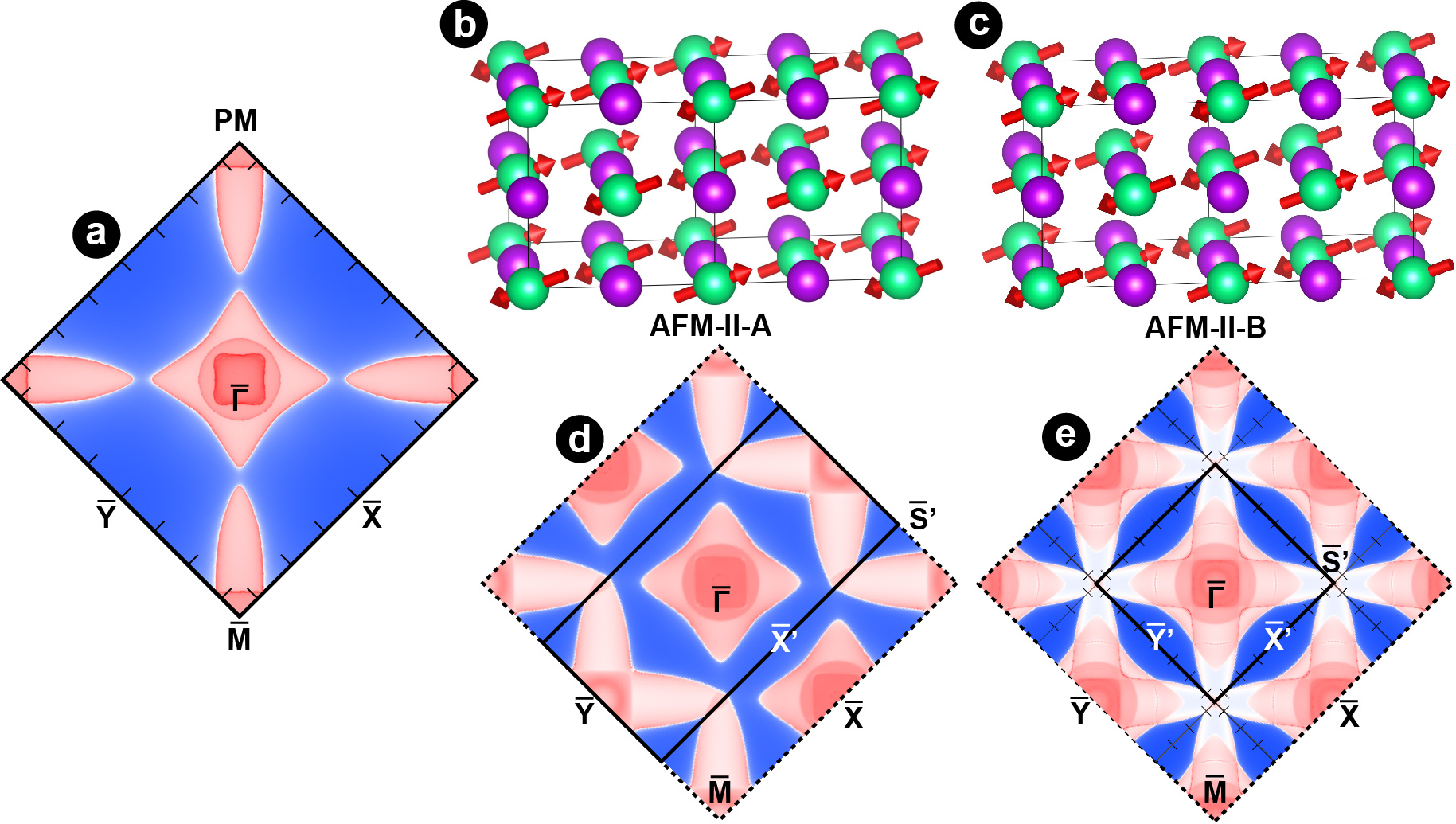}%
    \caption{Plot of calculated FS using DFT. 
    (a) 2D projections of bulk Fermi surface for PM phases. 
    (b) and (c) schematic representation of AFM-II-A and AFM-II-B order, respectively. 
    (d) and (e) 2D projections of bulk+surface Fermi surface for AFM-II-A and AFM-II-B order, respectively.
    In type AFM-II-A, the four face-centered cubic sublattices of antiparallel spins are correlated such that on each (111) plane, the spins are parallel, and the spins on two neighboring (111) planes are opposite. In type AFM-II-B, the four face-centered cubic sublattices divide the full face-centered cubic lattice in such a way that no two nearest neighboring ions are in the same sublattice. \cite{li1955magnetic}}
\end{figure}

In Fig.~1(c), we show a Fermi surface map of GdBi measured in the paramagnetic state. This Fermi surface is similar to the ones of other rare-earth monopnictides. It consists of several pockets around the center of the BZ, formed mostly by bulk states. This result agrees with DFT calculations [Fig.~2(a)]. Besides these pockets, the calculations predict the existence of two pockets formed by bulk states around the corner of the BZ. One of these pockets is partially seen in our experimental data. In the AFM state, we observe the development of additional states along $\overline{\Gamma}-\overline{X}$ direction. These states form a small sharp pocket (marked with green arrow) and a bigger broad star-shaped pocket (marked with red arrows) on the FS (Fig.~1(d)). Both of these pockets are centered around $\overline{X}$-point. The size and shape of the bigger pocket resemble the outer bulk pocket located around the $\overline{\Gamma}$-point. The comparison of the cuts taken along $\overline{\Gamma}-\overline{X}$ from PM and AFM data sets (Fig.~1(j, h)) show that the additional small pocket is formed by linear dispersion. The bands located near $\overline{\Gamma}$-point do not show any changes upon the magnetic transition. The cuts taken along $\overline{\Gamma}-\overline{M}$ direction (Fig.~1(g, j)) also show no qualitative change upon the magnetic transition. In Fig.~1(e, f), we show Fermi surface maps measured from DyBi in the PM and AFM states, respectively. These maps also demonstrate the development of additional states near the $\overline{X}$-point in the AFM state.

\begin{figure*}[t]
    \includegraphics[width=5 in]{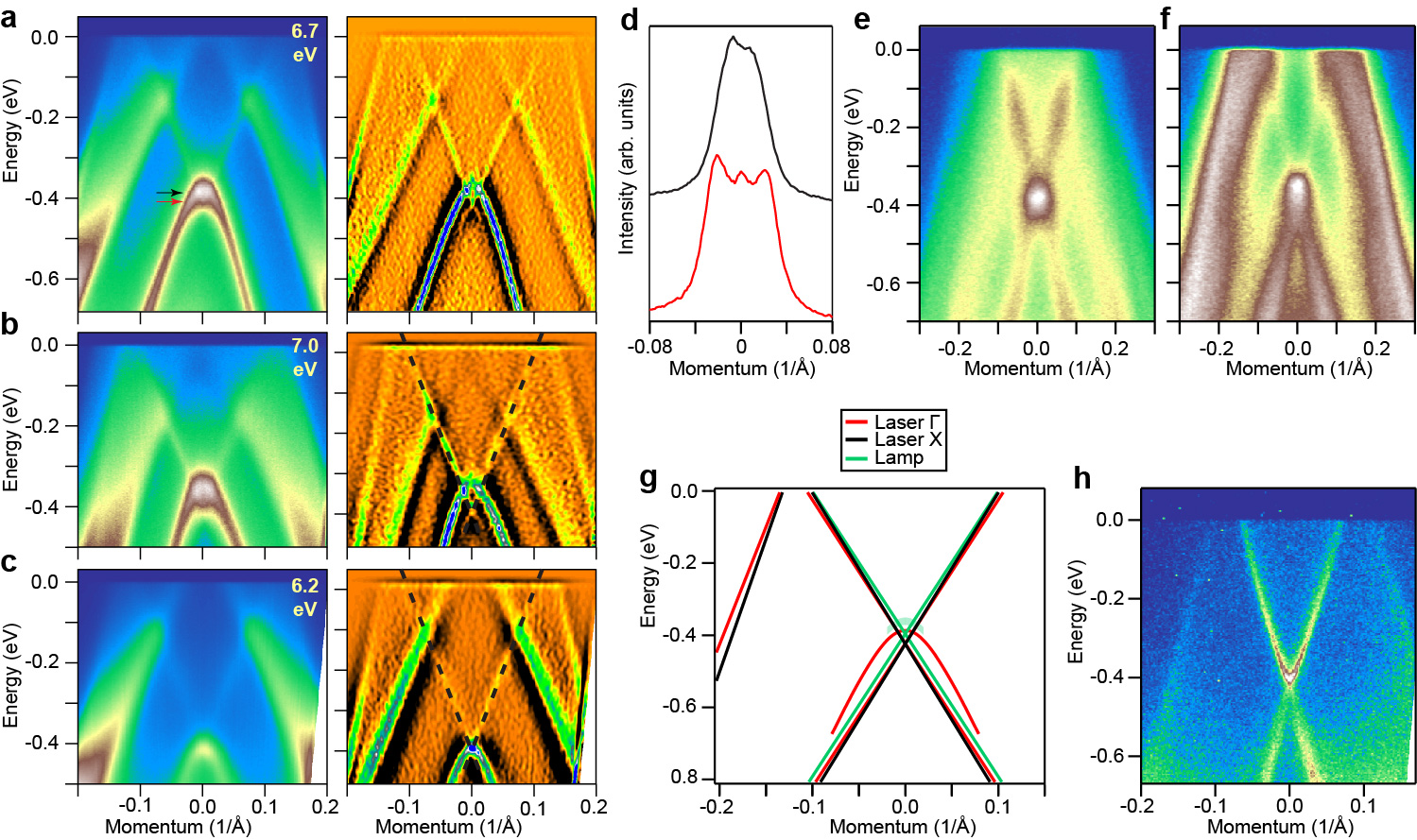}
    \caption{Band dispersion in AFM state of GdBi. 
    (a)-(c) Band dispersion measured along $\overline{\Gamma}-\overline{Y}$ direction (marked as \#3 in Fig.~1(d))
    using 6.7, 7, and 6.2~eV photons (left) and corresponding second derivative plots (right). These three spectra were measured at T~=~6.5, 4.5, 5.5~K, respectively. 
    (d) MDCs taken from (a) at 382 and 408~meV binding energy (marked with arrows of corresponding colors). 
    (e) and (f) Band dispersion measured along $\overline{\Gamma}-\overline{Y}$ direction in the first and the second BZ using 21.2 eV light from He-lamp at T~=~6.5~K.
    (g) Band dispersions extracted from plots (a),(d)-(f) by fitting MDCs. Points at zero-momentum were extracted by fitting EDCs. 
    (h) Band dispersion measured at T~=~6.5~K along the $\overline{X}-\overline{M}$ direction (marked as \#4 in Fig~1.(d)).}
\end{figure*}

These results are different from other rare-earth monopnictides \cite{kushnirenko2022rare,SchrunkNature2022,honma2023antiferromagnetic,kushnirenko2023directional,honma2023unusual,kushnirenko2024}. For comparison, we demonstrate a dataset measured from NdBi in AFM state in Fig.~1(b). These data do not indicate the presence of any states around the $\overline{X}$-point but instead show the development of Fermi arcs and elliptical pockets along $\overline{\Gamma}-\overline{M}$ direction.

The spectrum in Fig.~3(a) is measured in the AFM phase along $\overline{\Gamma}-\overline{Y}$ direction (cut \#3 marked in Fig.~1(d)). It demonstrates bands that, at first glance, resembles a Dirac cone. However, a closer inspection reveals that its upper part, instead of merging at one point with the lower part, crosses it, and the top of the lower band is rounded. The crossing can be even better seen in the second derivative plot and MDCs in Fig.~3(d). The MDC extracted through the top of the hole-like band(E~=~-382 meV) has two small peaks on the background of a large, broad peak. The additional peaks correspond to the upper band. The MDC extracted at E~=~-408 meV (below the top of the holelike band) has three peaks present. The outer two peaks correspond to the hole-like dispersion, and the central peak corresponds to the bottom of the upper band. The upper band is linear and less intense than hole-like dispersion. Thus, we can assume that they are parts of the Dirac cone, the lower part of which cannot be properly distinguished because of the strong background of the hole-like pocket. Only some hints of the lower part of the Dirac cone can be seen on the second derivative plot. To demonstrate the SS nature of the linear dispersions, we measured spectra similar to the one in Fig.~3(a) using light with different photon energy (see Fig.~3(b,c)). The linear dispersion shows no changes with photon energy. We fitted the shape of linear dispersions in the 6.7~eV spectrum and plotted the results over the second derivative plots of 6.2 and 7.0~eV spectra as dashed lines. The curves perfectly match the shape of the linear dispersions in these spectra. At the same time, the hole-like dispersion shows a substantial $k_z$-dependence: its top shifts by 65 meV between 6.2~and 7.0~eV spectra, and thus, it is definitely a bulk band, while the linear dispersion has SS origin.

For a better understanding of the electronic structure and, in particular, to prove the existence of the lower part of the Dirac cone, we performed measurements using a He-lamp ($h\nu$~=~21.2~eV). Fig.~3(e, f) demonstrate the spectra measured in $\overline{\Gamma}-\overline{Y}$ direction in the first and the second BZ, respectively. The lower part of the Dirac cone can be seen in both spectra. It is better seen in Fig.~3(e): here, we see two linear dispersions that cross at E~=~-406 meV and a blob of intensity above them (E~=~-356 meV), which corresponds to the top of the holelike bulk band. For further analysis, we extracted band dispersion from spectra in Fig.~3(a, e, f) by fitting the experimental data (see Fig.~6 in the appendix for details). The results of these fits are shown in Fig.~3(g). The linear dispersions extracted from 6.7 and 21.2~eV spectra match each other very well, which again proves that they are Dirac surface states.

\begin{figure}[t]
    \includegraphics[width=1\linewidth]{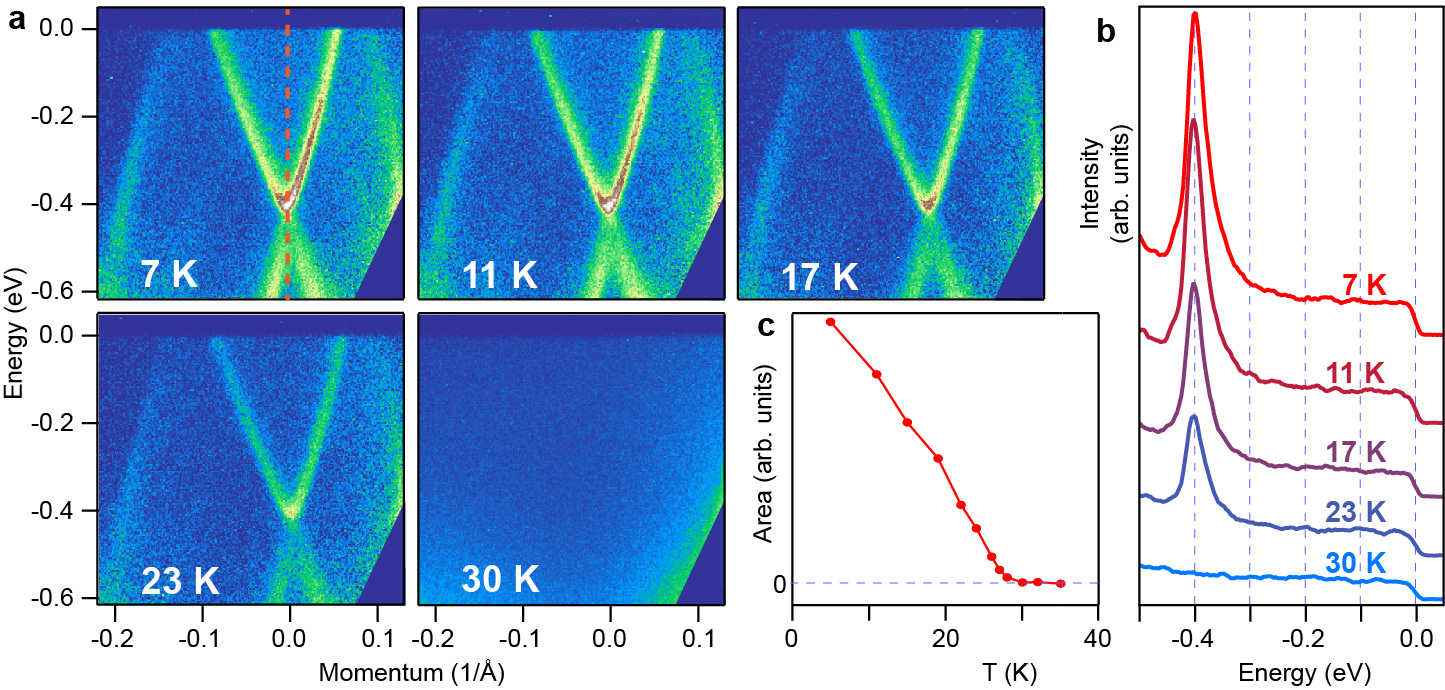}%
    \caption{Temperature dependence of the electronic structure of GdBi. 
    (a) Spectra measured along the $\overline{X}-\overline{M}$ direction at several temperatures. 
    (b) EDCs taken from data in panel (a) at zero-momentum (marked with dashed line). 
    (c) Temperature dependence of the area under the main peak from the EDCs in (b) after subtracting the area at 30 K.}
\end{figure}

In Fig.~3(h), we show a spectrum measured through $\overline{X}$-point (cut \#4 in Fig~1.(d)) in the AFM state. This spectrum has a sharp Dirac cone on the background of broad features. Strictly speaking, this spectrum is measured not along a straight cut for a constant value of $k_x$ component of momentum because of the geometry of the ARPES experiment. Because of this, in order to extract the real shape of the bands near the $\overline{X}$-point, several spectra should be analyzed (see SM for details). The shape of the upper part of the cone (cut through its center) extracted this way is shown in Fig.~3(g) with a black line. This shape matches the shape of the Dirac cone at the $\overline{\Gamma}$-point very well. This indicates that the bands that appear near the $\overline{X}$-point in the AFM state are the result of the folding of the bands from the center of the BZ. Another consequence of measuring at high angles of emission is that the photoemission experiment is expected to become less sensitive to the bulk states. Thus, the SS should be more pronounced. And indeed, the Diraclike feature folded from the $\overline{\Gamma}$-point becomes dominant near the $\overline{X}$-point. This is additional evidence that this Dirac-like feature at the $\overline{X}$-point and, as a consequence, the Dirac-like feature at the $\overline{\Gamma}$-point are originating from SS. Despite low bulk sensitivity, some bulk features can still be identified in Fig.~3(h). We have extracted the shape of the band that crosses the Fermi level at $\sim$0.13~$\AA^{-1}$ and compared it with the shape of the outermost band in Fig.~3(a). The results are shown in Fig.~3(g). These two band dispersions match each other reasonably well, which again demonstrates that the bands the $\overline{X}$-point are the result of band folding. 
The small difference between bulk bands in Fig.~3(a, h) can be explained by the $k_z$ dispersion of this bulk band ($k_z$ component also depends on the angle of emission).

\begin{figure}[b]
    \includegraphics[width=1\linewidth]{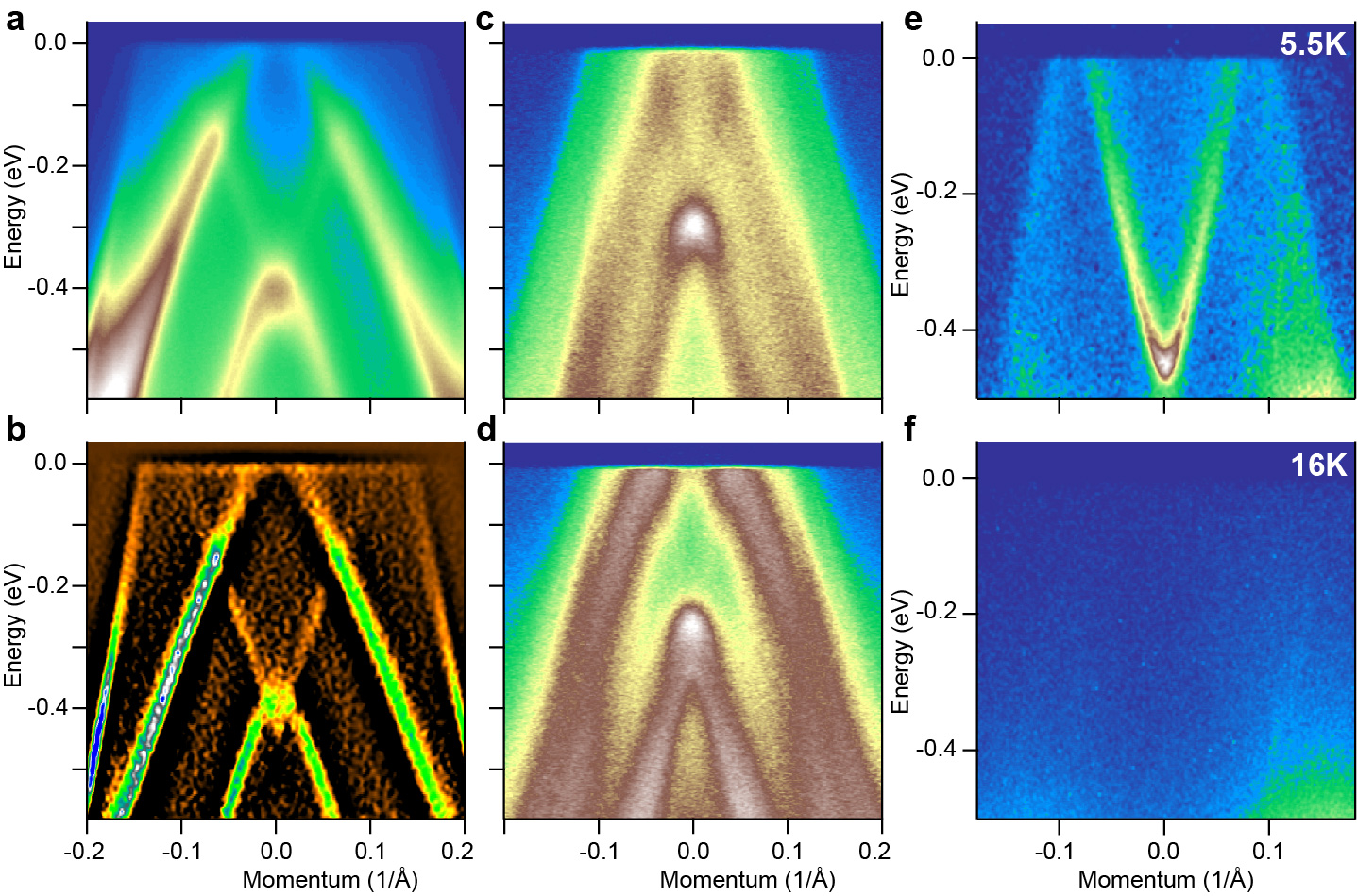}%
    \caption{Band dispersion of DyBi. 
    (a) and (b) Spectrum measured along $\overline{\Gamma}-\overline{Y}$ direction at T~=~16~K and the corresponding second derivative plot. 
    (c) and (d) Band dispersion measured T~=~6.6~K through the first and second $\overline{\Gamma}$-point using 21.2~eV light from He-lamp. 
    (e) and (f) Spectrum measured near the $\overline{X}$-point in AFM and PM state, respectively.}
\end{figure}

To analyze the evolution of the electronic structure near the $\overline{X}$-point with temperature, we measured a series of spectra through the $\overline{X}$-point at several temperatures below and above $T_N$. Five of these spectra are shown in Fig.~4(a). All spectra measured in the AFM state have the same set of bands. These bands do not change their shape or location with temperature. This can be better seen in Fig.~4(b), which demonstrates EDCs taken through the band center. The peak which corresponds to the Dirac point does not sift with temperature. However, the intensity of all bands at these spectra tends to decrease with increasing temperature, and all of them disappear above $T_N$ in the PM state. In Fig.~4(c), we plot a temperature dependence of the area under the EDC peak after subtracting the area above $T_N$. This plot demonstrates that the intensity of the bands folded to the area near the $\overline{X}$-point follows the AFM order parameter.

The folding of the bands from the center of the BZ to the region near the $\overline{X}$-point is predicted by the DFT calculations for the case of AFM-II-A order [Fig.~2(d)] agrees with our experimental results. Based on our data, we can exclude the scenario of AFM-II-B ordering. In this case, the folding of the bands from the $\overline{M}$-point to the center of the BZ is expected (Fig.~2(e)). This contradicts our experimental results, which show no changes at the $\overline{\Gamma}$-point upon AFM transition (see Fig.~1(g-j) and Fig.~7 in the appendix).

As was already mentioned, the FS of DyBi in the AFM phase (Fig.~1(f)) is similar to the one of GdBi (Fig.~1(d)). For further analysis, we show the Spectrum measured along $\overline{\Gamma}-\overline{Y}$ direction (Fig.~5(a)) and the corresponding second derivative plot (Fig.~5(b)). This spectrum has the same set of bands as Fig.~4(a-c), including a cone-like feature that crosses a more intense hole-like band. The spectra measured using 21.2~eV photons along $\overline{\Gamma}-\overline{Y}$ direction in the first and the second BZ are shown in Fig.~5(c, d), respectively. These spectra are similar to the spectra measured from GdBi. The separation between the lower part of the cone and the blob of intensity, which we associate with the top of the bulk hole-like dispersion, is more clearly seen in DyBi than in GdBi.

Fig.~5(e, f) demonstrate spectra measured near the $\overline{X}$-point in the AFM and PM state, respectively. In this experiment, we could not reach the $\overline{X}$-point where the upper and the lower parts of the cone are expected to touch each other. Nevertheless, in Fig.~5(e), one can see the upper parts of the cone and broad bulk features that are absent in the PM phase. This indicates that, in DyBi the bands from the center of the BZ are folded to the X-point in the AFM state in the same way as it happens in GdBi.

\section{CONCLUSIONS}
Our ARPES measurements demonstrate the emergence of new surface states near the $\overline{X}$-point in GdBi and DyBi upon a transition to AFM phase. These states are different from those observed in rare-earth monopnictides with AFM-I order (NdSb, NdBi, and CeBi).
The observed additional states are likely the result of the band folding from the center of the BZ. These states include a Dirac cone, which was observed in both $\overline{\Gamma}$- and $\overline{X}$-points of the BZ. The electronic structure near the center of the BZ does not change upon AFM transition. The observed results are consistent with AFM-II-A ordering in these materials.

\begin{figure}[t]
    \includegraphics[width=0.85\linewidth]{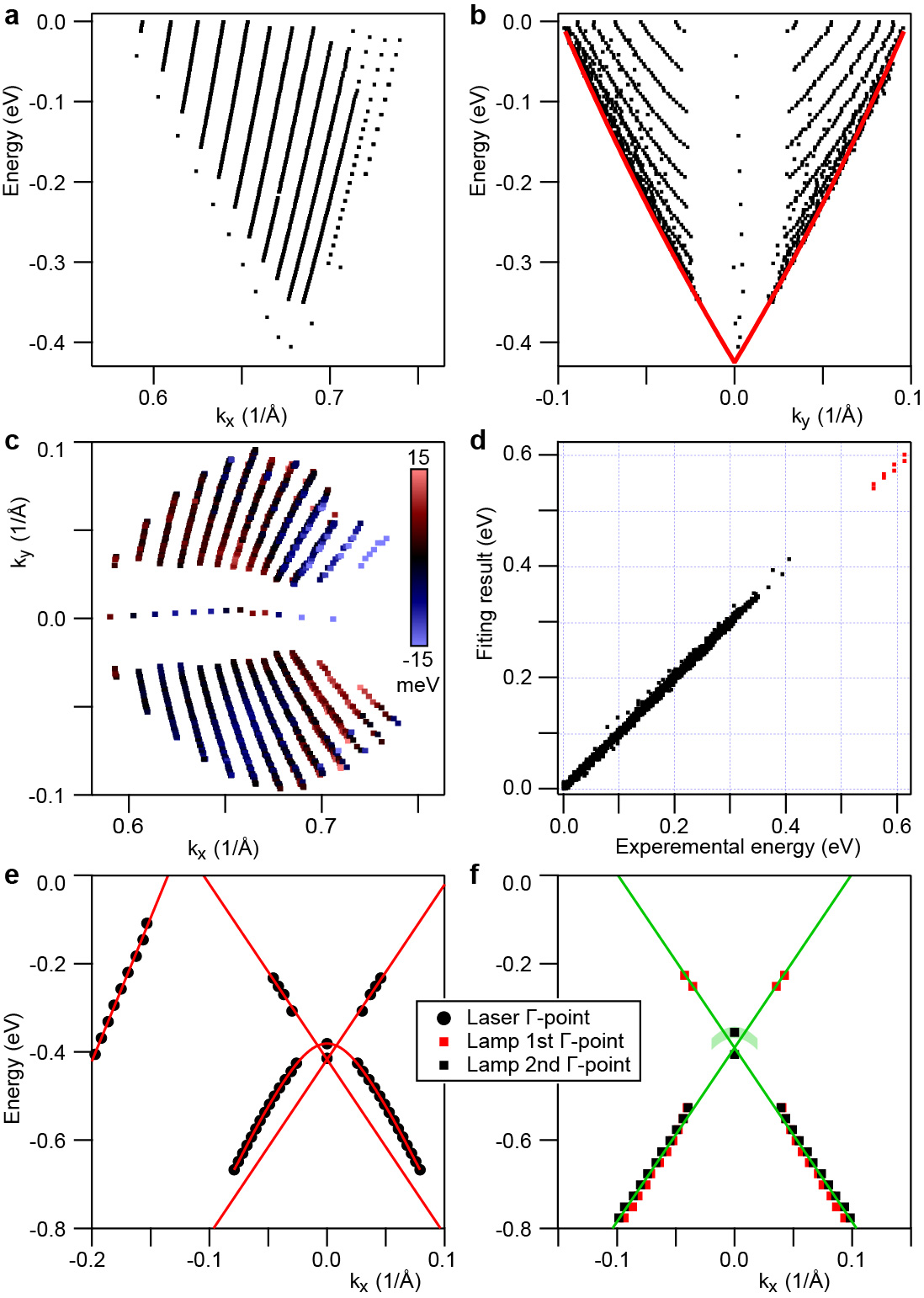}%
    \caption{Supplementary figure. 
    (a-c) tree projections of a set of points extracted by fitting 13 spectra measured near the $\overline{X}$-point. The red curve shows the share of the band dispersion along the $\overline{X}-\overline{M}$ cut extracted by fitting this set of points. 
    (d) correspondence between the experimental data and the results of the fit. 
    (e) set of points extracted by EDC and MDC fitting of the spectrum in Fig.~3(a). 
    (f) set of points extracted by EDC and MDC fitting of the spectra in Fig.~3(e,f).}
\end{figure}

\appendix
\section{Band shape fitting ($\overline{X}$-point)}
The measurement of the band structure near the $\overline{X}$-point with the 6.7~eV laser has to be performed at large take-off angles of the photoelectron: $\Theta\sim$ 71~degrees to reach the Dirac point. At such angles, the fact that the $k_y$ is not equal in all parts of the spectrum cannot be neglected. Thus, in order to extract the shape of the bands along a straight cut at a constant value of $k_y$, we need to compile data from several closely located cuts.

The shape of the Dirac cone near the $\overline{X}$-point was extracted by fitting 13 spectra measured at angles from 51 to 75~degrees. The set of points expected by fitting MDC peaks is shown in Fig.~6(a-c). In the next step, we fit this set of points with the following function: $E=E_0+A_1R+A_2R+A_3R^3$, where $R=\sqrt{(k_x-k_{x0})^2+(k_y-k_{y0})^2}(1+Bcos(4\Theta))$, and $\Theta=arctan(k_x-k_{x0})/(k_y-k_{y0}))$. Here $Bcos(4\Theta)$ component represents a deviation of the pocket shape from a perfect circle. The fit gives the following values of the coefficients: $E_0=$~245$\pm$3; $k_{x0}=$~0.6809$\pm$0.0001; $k_{y0}=$~0.008$\pm$0.0001; $A_1=$~3.87$\pm$0.18; $A_2=$~5.4$\pm$3.4; $A_3=$~19$\pm$19;  $B=$~-0.046$\pm$0.011.

Fig.~6(d), which shows a correspondence between the experimental data and the results of the fit, and the colors in Fig.~6(c), which show a difference between the experimental data and the results of the fit for different parts of the band, indicate good quality of the fit. If we extend the function, which describes the band shape, to the lower cone by changing the sign of effective radius, the result would also match the experimental data in that energy range (see red markers in Fig.~6(d)). The fit coefficients for non-linear components of the fit ($A_2R^2$ and $A_3R^3$) are comparable to the fit error. Also, these components are insignificant, which can be visually seen from the shape of the red curve in Fig.~6(b), which represents the results of the fit at $k_x$ = $k_{x0}$. All this indicates that the observed band is indeed a Dirac cone.

\section{Band shape fitting ($\overline{\Gamma}$-point)}
In Fig.~6(e, f), we demonstrate the intermediate steps of extracting the shape of the bands in Fig.~2(a,d, e). The markers at $k_x\neq0$ represent the results of fitting MDC peaks, and the markers $k_x=0$ represent the results of fitting EDC peaks. The lines represent the results of fitting these points. The resulting shape of the upper band from the spectrum measured with the laser was fitted with a line, and the shape of the lower band was fitted with a polynomial of degree four. The shape of the band from the spectra measured with the He-lamp was fitted with a line.

\begin{figure*}[t]
    \includegraphics[width=6.5 in]{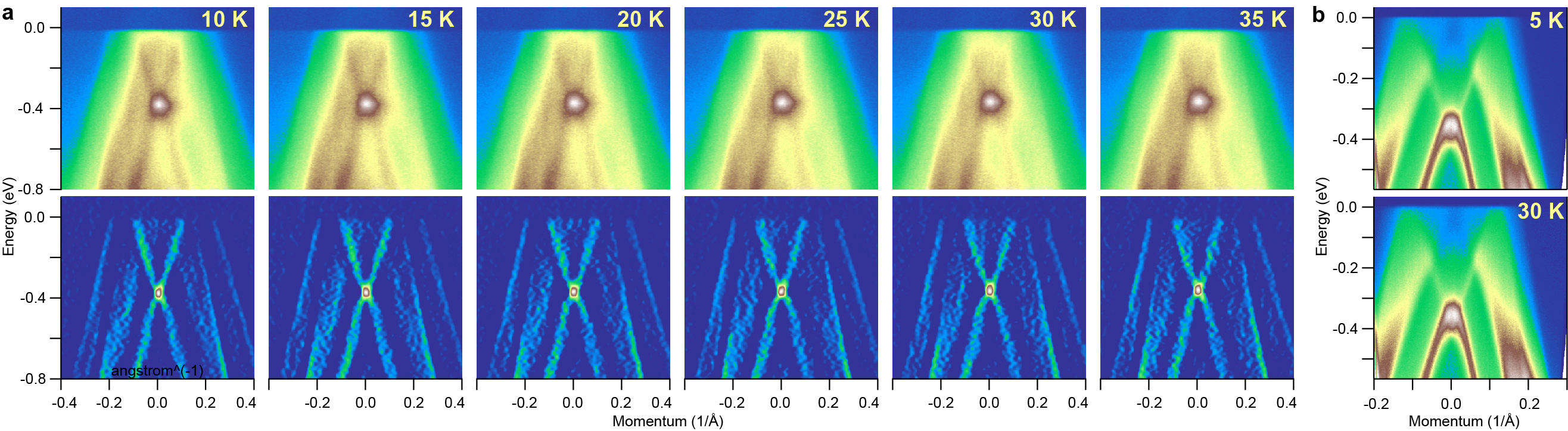}
    \caption{GdBi temperature dependence: additional data. 
    (a) Spectra measured along $\overline{\Gamma}-\overline{Y}$ direction using 21.2 eV light (top) and corresponding second derivative plots (bottom). 
    (b) Spectra measured along $\overline{\Gamma}-\overline{Y}$ direction using 7.0 eV light.}
\end{figure*}

\bibliography{ndBi_arcs}

\begin{thebibliography}{39}%
\makeatletter
\providecommand \@ifxundefined [1]{%
 \@ifx{#1\undefined}
}%
\providecommand \@ifnum [1]{%
 \ifnum #1\expandafter \@firstoftwo
 \else \expandafter \@secondoftwo
 \fi
}%
\providecommand \@ifx [1]{%
 \ifx #1\expandafter \@firstoftwo
 \else \expandafter \@secondoftwo
 \fi
}%
\providecommand \natexlab [1]{#1}%
\providecommand \enquote  [1]{``#1''}%
\providecommand \bibnamefont  [1]{#1}%
\providecommand \bibfnamefont [1]{#1}%
\providecommand \citenamefont [1]{#1}%
\providecommand \href@noop [0]{\@secondoftwo}%
\providecommand \href [0]{\begingroup \@sanitize@url \@href}%
\providecommand \@href[1]{\@@startlink{#1}\@@href}%
\providecommand \@@href[1]{\endgroup#1\@@endlink}%
\providecommand \@sanitize@url [0]{\catcode `\\12\catcode `\$12\catcode
  `\&12\catcode `\#12\catcode `\^12\catcode `\_12\catcode `\%12\relax}%
\providecommand \@@startlink[1]{}%
\providecommand \@@endlink[0]{}%
\providecommand \url  [0]{\begingroup\@sanitize@url \@url }%
\providecommand \@url [1]{\endgroup\@href {#1}{\urlprefix }}%
\providecommand \urlprefix  [0]{URL }%
\providecommand \Eprint [0]{\href }%
\providecommand \doibase [0]{http://dx.doi.org/}%
\providecommand \selectlanguage [0]{\@gobble}%
\providecommand \bibinfo  [0]{\@secondoftwo}%
\providecommand \bibfield  [0]{\@secondoftwo}%
\providecommand \translation [1]{[#1]}%
\providecommand \BibitemOpen [0]{}%
\providecommand \bibitemStop [0]{}%
\providecommand \bibitemNoStop [0]{.\EOS\space}%
\providecommand \EOS [0]{\spacefactor3000\relax}%
\providecommand \BibitemShut  [1]{\csname bibitem#1\endcsname}%
\let\auto@bib@innerbib\@empty
\bibitem [{\citenamefont {Kushnirenko}\ \emph {et~al.}(2022)\citenamefont
  {Kushnirenko}, \citenamefont {Schrunk}, \citenamefont {Kuthanazhi},
  \citenamefont {Wang}, \citenamefont {Ahn}, \citenamefont {O'Leary},
  \citenamefont {Eaton}, \citenamefont {Bud'ko}, \citenamefont {Slager},
  \citenamefont {Canfield},\ and\ \citenamefont
  {Kaminski}}]{kushnirenko2022rare}%
  \BibitemOpen
  \bibfield  {author} {\bibinfo {author} {\bibfnamefont {Y.}~\bibnamefont
  {Kushnirenko}}, \bibinfo {author} {\bibfnamefont {B.}~\bibnamefont
  {Schrunk}}, \bibinfo {author} {\bibfnamefont {B.}~\bibnamefont {Kuthanazhi}},
  \bibinfo {author} {\bibfnamefont {L.-L.}\ \bibnamefont {Wang}}, \bibinfo
  {author} {\bibfnamefont {J.}~\bibnamefont {Ahn}}, \bibinfo {author}
  {\bibfnamefont {E.}~\bibnamefont {O'Leary}}, \bibinfo {author} {\bibfnamefont
  {A.}~\bibnamefont {Eaton}}, \bibinfo {author} {\bibfnamefont {S.~L.}\
  \bibnamefont {Bud'ko}}, \bibinfo {author} {\bibfnamefont {R.-J.}\
  \bibnamefont {Slager}}, \bibinfo {author} {\bibfnamefont {P.~C.}\
  \bibnamefont {Canfield}}, \ and\ \bibinfo {author} {\bibfnamefont
  {A.}~\bibnamefont {Kaminski}},\ }\href {\doibase 10.1103/PhysRevB.106.115112}
  {\bibfield  {journal} {\bibinfo  {journal} {Phys. Rev. B}\ }\textbf {\bibinfo
  {volume} {106}},\ \bibinfo {pages} {115112} (\bibinfo {year}
  {2022})}\BibitemShut {NoStop}%
\bibitem [{\citenamefont {Schrunk}\ \emph {et~al.}(2022)\citenamefont
  {Schrunk}, \citenamefont {Kushnirenko}, \citenamefont {Kuthanazhi},
  \citenamefont {Ahn}, \citenamefont {Wang}, \citenamefont {O’Leary},
  \citenamefont {Lee}, \citenamefont {Eaton}, \citenamefont {Fedorov},
  \citenamefont {Lou}, \citenamefont {Voroshnin}, \citenamefont {Clark},
  \citenamefont {Sánchez-Barriga}, \citenamefont {Bud’ko}, \citenamefont
  {Slager}, \citenamefont {Canfield},\ and\ \citenamefont
  {Kaminski}}]{SchrunkNature2022}%
  \BibitemOpen
  \bibfield  {author} {\bibinfo {author} {\bibfnamefont {B.}~\bibnamefont
  {Schrunk}}, \bibinfo {author} {\bibfnamefont {Y.}~\bibnamefont
  {Kushnirenko}}, \bibinfo {author} {\bibfnamefont {B.}~\bibnamefont
  {Kuthanazhi}}, \bibinfo {author} {\bibfnamefont {J.}~\bibnamefont {Ahn}},
  \bibinfo {author} {\bibfnamefont {L.-L.}\ \bibnamefont {Wang}}, \bibinfo
  {author} {\bibfnamefont {E.}~\bibnamefont {O’Leary}}, \bibinfo {author}
  {\bibfnamefont {K.}~\bibnamefont {Lee}}, \bibinfo {author} {\bibfnamefont
  {A.}~\bibnamefont {Eaton}}, \bibinfo {author} {\bibfnamefont
  {A.}~\bibnamefont {Fedorov}}, \bibinfo {author} {\bibfnamefont
  {R.}~\bibnamefont {Lou}}, \bibinfo {author} {\bibfnamefont {V.}~\bibnamefont
  {Voroshnin}}, \bibinfo {author} {\bibfnamefont {O.~J.}\ \bibnamefont
  {Clark}}, \bibinfo {author} {\bibfnamefont {J.}~\bibnamefont
  {Sánchez-Barriga}}, \bibinfo {author} {\bibfnamefont {S.~L.}\ \bibnamefont
  {Bud’ko}}, \bibinfo {author} {\bibfnamefont {R.-J.}\ \bibnamefont
  {Slager}}, \bibinfo {author} {\bibfnamefont {P.~C.}\ \bibnamefont
  {Canfield}}, \ and\ \bibinfo {author} {\bibfnamefont {A.}~\bibnamefont
  {Kaminski}},\ }\href@noop {} {\bibfield  {journal} {\bibinfo  {journal}
  {Nature}\ }\textbf {\bibinfo {volume} {603}},\ \bibinfo {pages} {610–615}
  (\bibinfo {year} {2022})}\BibitemShut {NoStop}%
\bibitem [{\citenamefont {Honma}\ \emph
  {et~al.}(2023{\natexlab{a}})\citenamefont {Honma}, \citenamefont {Takane},
  \citenamefont {Souma}, \citenamefont {Yamauchi}, \citenamefont {Wang},
  \citenamefont {Nakayama}, \citenamefont {Sugawara}, \citenamefont {Kitamura},
  \citenamefont {Horiba}, \citenamefont {Kumigashira} \emph
  {et~al.}}]{honma2023antiferromagnetic}%
  \BibitemOpen
  \bibfield  {author} {\bibinfo {author} {\bibfnamefont {A.}~\bibnamefont
  {Honma}}, \bibinfo {author} {\bibfnamefont {D.}~\bibnamefont {Takane}},
  \bibinfo {author} {\bibfnamefont {S.}~\bibnamefont {Souma}}, \bibinfo
  {author} {\bibfnamefont {K.}~\bibnamefont {Yamauchi}}, \bibinfo {author}
  {\bibfnamefont {Y.}~\bibnamefont {Wang}}, \bibinfo {author} {\bibfnamefont
  {K.}~\bibnamefont {Nakayama}}, \bibinfo {author} {\bibfnamefont
  {K.}~\bibnamefont {Sugawara}}, \bibinfo {author} {\bibfnamefont
  {M.}~\bibnamefont {Kitamura}}, \bibinfo {author} {\bibfnamefont
  {K.}~\bibnamefont {Horiba}}, \bibinfo {author} {\bibfnamefont
  {H.}~\bibnamefont {Kumigashira}},  \emph {et~al.},\ }\href@noop {} {\bibfield
   {journal} {\bibinfo  {journal} {Nature Communications}\ }\textbf {\bibinfo
  {volume} {14}},\ \bibinfo {pages} {7396} (\bibinfo {year}
  {2023}{\natexlab{a}})}\BibitemShut {NoStop}%
\bibitem [{\citenamefont {Kushnirenko}\ \emph {et~al.}(2023)\citenamefont
  {Kushnirenko}, \citenamefont {Kuthanazhi}, \citenamefont {Wang},
  \citenamefont {Schrunk}, \citenamefont {O'Leary}, \citenamefont {Eaton},
  \citenamefont {Canfield},\ and\ \citenamefont
  {Kaminski}}]{kushnirenko2023directional}%
  \BibitemOpen
  \bibfield  {author} {\bibinfo {author} {\bibfnamefont {Y.}~\bibnamefont
  {Kushnirenko}}, \bibinfo {author} {\bibfnamefont {B.}~\bibnamefont
  {Kuthanazhi}}, \bibinfo {author} {\bibfnamefont {L.-L.}\ \bibnamefont
  {Wang}}, \bibinfo {author} {\bibfnamefont {B.}~\bibnamefont {Schrunk}},
  \bibinfo {author} {\bibfnamefont {E.}~\bibnamefont {O'Leary}}, \bibinfo
  {author} {\bibfnamefont {A.}~\bibnamefont {Eaton}}, \bibinfo {author}
  {\bibfnamefont {P.~C.}\ \bibnamefont {Canfield}}, \ and\ \bibinfo {author}
  {\bibfnamefont {A.}~\bibnamefont {Kaminski}},\ }\href@noop {} {\bibfield
  {journal} {\bibinfo  {journal} {Physical Review B}\ }\textbf {\bibinfo
  {volume} {108}},\ \bibinfo {pages} {115102} (\bibinfo {year}
  {2023})}\BibitemShut {NoStop}%
\bibitem [{\citenamefont {Honma}\ \emph
  {et~al.}(2023{\natexlab{b}})\citenamefont {Honma}, \citenamefont {Takane},
  \citenamefont {Souma}, \citenamefont {Wang}, \citenamefont {Nakayama},
  \citenamefont {Kitamura}, \citenamefont {Horiba}, \citenamefont
  {Kumigashira}, \citenamefont {Takahashi}, \citenamefont {Ando} \emph
  {et~al.}}]{honma2023unusual}%
  \BibitemOpen
  \bibfield  {author} {\bibinfo {author} {\bibfnamefont {A.}~\bibnamefont
  {Honma}}, \bibinfo {author} {\bibfnamefont {D.}~\bibnamefont {Takane}},
  \bibinfo {author} {\bibfnamefont {S.}~\bibnamefont {Souma}}, \bibinfo
  {author} {\bibfnamefont {Y.}~\bibnamefont {Wang}}, \bibinfo {author}
  {\bibfnamefont {K.}~\bibnamefont {Nakayama}}, \bibinfo {author}
  {\bibfnamefont {M.}~\bibnamefont {Kitamura}}, \bibinfo {author}
  {\bibfnamefont {K.}~\bibnamefont {Horiba}}, \bibinfo {author} {\bibfnamefont
  {H.}~\bibnamefont {Kumigashira}}, \bibinfo {author} {\bibfnamefont
  {T.}~\bibnamefont {Takahashi}}, \bibinfo {author} {\bibfnamefont
  {Y.}~\bibnamefont {Ando}},  \emph {et~al.},\ }\href@noop {} {\bibfield
  {journal} {\bibinfo  {journal} {Physical Review B}\ }\textbf {\bibinfo
  {volume} {108}},\ \bibinfo {pages} {115118} (\bibinfo {year}
  {2023}{\natexlab{b}})}\BibitemShut {NoStop}%
\bibitem [{\citenamefont {Tsuchida}\ and\ \citenamefont
  {Wallace}(1965)}]{Tsuchida_1965}%
  \BibitemOpen
  \bibfield  {author} {\bibinfo {author} {\bibfnamefont {T.}~\bibnamefont
  {Tsuchida}}\ and\ \bibinfo {author} {\bibfnamefont {W.~E.}\ \bibnamefont
  {Wallace}},\ }\href@noop {} {\bibfield  {journal} {\bibinfo  {journal} {The
  Journal of Chemical Physics}\ }\textbf {\bibinfo {volume} {43}},\ \bibinfo
  {pages} {2087} (\bibinfo {year} {1965})}\BibitemShut {NoStop}%
\bibitem [{\citenamefont {Nereson}\ and\ \citenamefont
  {Arnold}(1971)}]{nereson_1971}%
  \BibitemOpen
  \bibfield  {author} {\bibinfo {author} {\bibfnamefont {N.}~\bibnamefont
  {Nereson}}\ and\ \bibinfo {author} {\bibfnamefont {G.}~\bibnamefont
  {Arnold}},\ }\href@noop {} {\bibfield  {journal} {\bibinfo  {journal}
  {Journal of Applied Physics}\ }\textbf {\bibinfo {volume} {42}},\ \bibinfo
  {pages} {1625} (\bibinfo {year} {1971})}\BibitemShut {NoStop}%
\bibitem [{\citenamefont {Bartholin}\ \emph {et~al.}(1979)\citenamefont
  {Bartholin}, \citenamefont {Burlet}, \citenamefont {Quezel}, \citenamefont
  {Rossat-Mignod},\ and\ \citenamefont {Vogt}}]{bartholin1979hydrostatic}%
  \BibitemOpen
  \bibfield  {author} {\bibinfo {author} {\bibfnamefont {H.}~\bibnamefont
  {Bartholin}}, \bibinfo {author} {\bibfnamefont {P.}~\bibnamefont {Burlet}},
  \bibinfo {author} {\bibfnamefont {S.}~\bibnamefont {Quezel}}, \bibinfo
  {author} {\bibfnamefont {J.}~\bibnamefont {Rossat-Mignod}}, \ and\ \bibinfo
  {author} {\bibfnamefont {O.}~\bibnamefont {Vogt}},\ }\href@noop {} {\bibfield
   {journal} {\bibinfo  {journal} {Le Journal de Physique Colloques}\ }\textbf
  {\bibinfo {volume} {40}},\ \bibinfo {pages} {C5} (\bibinfo {year}
  {1979})}\BibitemShut {NoStop}%
\bibitem [{\citenamefont {Kumigashira}\ \emph {et~al.}(1996)\citenamefont
  {Kumigashira}, \citenamefont {Yang}, \citenamefont {Yokoya}, \citenamefont
  {Chainani}, \citenamefont {Takahashi}, \citenamefont {Uesawa}, \citenamefont
  {Suzuki}, \citenamefont {Sakai},\ and\ \citenamefont
  {Kaneta}}]{KumigashiraPRB1996}%
  \BibitemOpen
  \bibfield  {author} {\bibinfo {author} {\bibfnamefont {H.}~\bibnamefont
  {Kumigashira}}, \bibinfo {author} {\bibfnamefont {S.-H.}\ \bibnamefont
  {Yang}}, \bibinfo {author} {\bibfnamefont {T.}~\bibnamefont {Yokoya}},
  \bibinfo {author} {\bibfnamefont {A.}~\bibnamefont {Chainani}}, \bibinfo
  {author} {\bibfnamefont {T.}~\bibnamefont {Takahashi}}, \bibinfo {author}
  {\bibfnamefont {A.}~\bibnamefont {Uesawa}}, \bibinfo {author} {\bibfnamefont
  {T.}~\bibnamefont {Suzuki}}, \bibinfo {author} {\bibfnamefont
  {O.}~\bibnamefont {Sakai}}, \ and\ \bibinfo {author} {\bibfnamefont
  {Y.}~\bibnamefont {Kaneta}},\ }\href {\doibase 10.1103/PhysRevB.54.9341}
  {\bibfield  {journal} {\bibinfo  {journal} {Phys. Rev. B}\ }\textbf {\bibinfo
  {volume} {54}},\ \bibinfo {pages} {9341} (\bibinfo {year}
  {1996})}\BibitemShut {NoStop}%
\bibitem [{\citenamefont {Rossat-Mignod}\ \emph {et~al.}(1983)\citenamefont
  {Rossat-Mignod}, \citenamefont {Burlet}, \citenamefont {Quezel},
  \citenamefont {Effantin}, \citenamefont {Delac{\^o}te}, \citenamefont
  {Bartholin}, \citenamefont {Vogt},\ and\ \citenamefont
  {Ravot}}]{rossat1983magnetic}%
  \BibitemOpen
  \bibfield  {author} {\bibinfo {author} {\bibfnamefont {J.}~\bibnamefont
  {Rossat-Mignod}}, \bibinfo {author} {\bibfnamefont {P.}~\bibnamefont
  {Burlet}}, \bibinfo {author} {\bibfnamefont {S.}~\bibnamefont {Quezel}},
  \bibinfo {author} {\bibfnamefont {J.}~\bibnamefont {Effantin}}, \bibinfo
  {author} {\bibfnamefont {D.}~\bibnamefont {Delac{\^o}te}}, \bibinfo {author}
  {\bibfnamefont {H.}~\bibnamefont {Bartholin}}, \bibinfo {author}
  {\bibfnamefont {O.}~\bibnamefont {Vogt}}, \ and\ \bibinfo {author}
  {\bibfnamefont {D.}~\bibnamefont {Ravot}},\ }\href@noop {} {\bibfield
  {journal} {\bibinfo  {journal} {Journal of Magnetism and Magnetic Materials}\
  }\textbf {\bibinfo {volume} {31}},\ \bibinfo {pages} {398} (\bibinfo {year}
  {1983})}\BibitemShut {NoStop}%
\bibitem [{\citenamefont {Wiener}\ and\ \citenamefont
  {Canfield}(2000)}]{wiener2000magnetic}%
  \BibitemOpen
  \bibfield  {author} {\bibinfo {author} {\bibfnamefont {T.}~\bibnamefont
  {Wiener}}\ and\ \bibinfo {author} {\bibfnamefont {P.}~\bibnamefont
  {Canfield}},\ }\href@noop {} {\bibfield  {journal} {\bibinfo  {journal}
  {Journal of Alloys and Compounds}\ }\textbf {\bibinfo {volume} {303}},\
  \bibinfo {pages} {505} (\bibinfo {year} {2000})}\BibitemShut {NoStop}%
\bibitem [{\citenamefont {Kuroda}\ \emph {et~al.}(2020)\citenamefont {Kuroda},
  \citenamefont {Arai}, \citenamefont {Rezaei}, \citenamefont {Kunisada},
  \citenamefont {Sakuragi}, \citenamefont {Alaei}, \citenamefont {Kinoshita},
  \citenamefont {Bareille}, \citenamefont {Noguchi}, \citenamefont {Nakayama},
  \citenamefont {Akebi}, \citenamefont {Sakano}, \citenamefont {Kawaguchi},
  \citenamefont {Arita}, \citenamefont {Ideta}, \citenamefont {Tanaka},
  \citenamefont {Kitazawa}, \citenamefont {Okazaki}, \citenamefont {Tokunaga},
  \citenamefont {Haga}, \citenamefont {Shin}, \citenamefont {Suzuki},
  \citenamefont {Arita},\ and\ \citenamefont {Kondo}}]{Kuroda_2020}%
  \BibitemOpen
  \bibfield  {author} {\bibinfo {author} {\bibfnamefont {K.}~\bibnamefont
  {Kuroda}}, \bibinfo {author} {\bibfnamefont {Y.}~\bibnamefont {Arai}},
  \bibinfo {author} {\bibfnamefont {N.}~\bibnamefont {Rezaei}}, \bibinfo
  {author} {\bibfnamefont {S.}~\bibnamefont {Kunisada}}, \bibinfo {author}
  {\bibfnamefont {S.}~\bibnamefont {Sakuragi}}, \bibinfo {author}
  {\bibfnamefont {M.}~\bibnamefont {Alaei}}, \bibinfo {author} {\bibfnamefont
  {Y.}~\bibnamefont {Kinoshita}}, \bibinfo {author} {\bibfnamefont
  {C.}~\bibnamefont {Bareille}}, \bibinfo {author} {\bibfnamefont
  {R.}~\bibnamefont {Noguchi}}, \bibinfo {author} {\bibfnamefont
  {M.}~\bibnamefont {Nakayama}}, \bibinfo {author} {\bibfnamefont
  {S.}~\bibnamefont {Akebi}}, \bibinfo {author} {\bibfnamefont
  {M.}~\bibnamefont {Sakano}}, \bibinfo {author} {\bibfnamefont
  {K.}~\bibnamefont {Kawaguchi}}, \bibinfo {author} {\bibfnamefont
  {M.}~\bibnamefont {Arita}}, \bibinfo {author} {\bibfnamefont
  {S.}~\bibnamefont {Ideta}}, \bibinfo {author} {\bibfnamefont
  {K.}~\bibnamefont {Tanaka}}, \bibinfo {author} {\bibfnamefont
  {H.}~\bibnamefont {Kitazawa}}, \bibinfo {author} {\bibfnamefont
  {K.}~\bibnamefont {Okazaki}}, \bibinfo {author} {\bibfnamefont
  {M.}~\bibnamefont {Tokunaga}}, \bibinfo {author} {\bibfnamefont
  {Y.}~\bibnamefont {Haga}}, \bibinfo {author} {\bibfnamefont {S.}~\bibnamefont
  {Shin}}, \bibinfo {author} {\bibfnamefont {H.~S.}\ \bibnamefont {Suzuki}},
  \bibinfo {author} {\bibfnamefont {R.}~\bibnamefont {Arita}}, \ and\ \bibinfo
  {author} {\bibfnamefont {T.}~\bibnamefont {Kondo}},\ }\href@noop {}
  {\bibfield  {journal} {\bibinfo  {journal} {Nature Communications}\ }\textbf
  {\bibinfo {volume} {11}},\ \bibinfo {pages} {1} (\bibinfo {year}
  {2020})}\BibitemShut {NoStop}%
\bibitem [{\citenamefont {Kuthanazhi}\ \emph {et~al.}(2022)\citenamefont
  {Kuthanazhi}, \citenamefont {Jo}, \citenamefont {Xiang}, \citenamefont
  {Bud'ko},\ and\ \citenamefont {Canfield}}]{kuthanazhi2022magnetisation}%
  \BibitemOpen
  \bibfield  {author} {\bibinfo {author} {\bibfnamefont {B.}~\bibnamefont
  {Kuthanazhi}}, \bibinfo {author} {\bibfnamefont {N.~H.}\ \bibnamefont {Jo}},
  \bibinfo {author} {\bibfnamefont {L.}~\bibnamefont {Xiang}}, \bibinfo
  {author} {\bibfnamefont {S.~L.}\ \bibnamefont {Bud'ko}}, \ and\ \bibinfo
  {author} {\bibfnamefont {P.~C.}\ \bibnamefont {Canfield}},\ }\href@noop {}
  {\bibfield  {journal} {\bibinfo  {journal} {Philosophical Magazine}\ }\textbf
  {\bibinfo {volume} {102}},\ \bibinfo {pages} {542} (\bibinfo {year}
  {2022})}\BibitemShut {NoStop}%
\bibitem [{\citenamefont {Kushnirenko}(2024)}]{kushnirenko2024}%
  \BibitemOpen
  \bibfield  {author} {\bibinfo {author} {\bibfnamefont {Y.}~\bibnamefont
  {Kushnirenko}},\ }\href@noop {} {\bibfield  {journal} {\bibinfo  {journal}
  {Unpublished}\ } (\bibinfo {year} {2024})}\BibitemShut {NoStop}%
\bibitem [{\citenamefont {Guo}\ \emph {et~al.}(2017)\citenamefont {Guo},
  \citenamefont {Cao}, \citenamefont {Smidman}, \citenamefont {Wu},
  \citenamefont {Zhang}, \citenamefont {Steglich}, \citenamefont {Zhang},\ and\
  \citenamefont {Yuan}}]{GuoNPJ2017}%
  \BibitemOpen
  \bibfield  {author} {\bibinfo {author} {\bibfnamefont {C.}~\bibnamefont
  {Guo}}, \bibinfo {author} {\bibfnamefont {C.}~\bibnamefont {Cao}}, \bibinfo
  {author} {\bibfnamefont {M.}~\bibnamefont {Smidman}}, \bibinfo {author}
  {\bibfnamefont {F.}~\bibnamefont {Wu}}, \bibinfo {author} {\bibfnamefont
  {Y.}~\bibnamefont {Zhang}}, \bibinfo {author} {\bibfnamefont
  {F.}~\bibnamefont {Steglich}}, \bibinfo {author} {\bibfnamefont {F.-C.}\
  \bibnamefont {Zhang}}, \ and\ \bibinfo {author} {\bibfnamefont
  {H.}~\bibnamefont {Yuan}},\ }\href@noop {} {\bibfield  {journal} {\bibinfo
  {journal} {Npj Quantum Materials}\ }\textbf {\bibinfo {volume} {2}},\
  (\bibinfo {year} {2017})}\BibitemShut {NoStop}%
\bibitem [{\citenamefont {Li}\ \emph {et~al.}(2017)\citenamefont {Li},
  \citenamefont {Xu}, \citenamefont {Ning}, \citenamefont {Su}, \citenamefont
  {Iitaka}, \citenamefont {Tohyama},\ and\ \citenamefont
  {Zhang}}]{li2017predicted}%
  \BibitemOpen
  \bibfield  {author} {\bibinfo {author} {\bibfnamefont {Z.}~\bibnamefont
  {Li}}, \bibinfo {author} {\bibfnamefont {D.-D.}\ \bibnamefont {Xu}}, \bibinfo
  {author} {\bibfnamefont {S.-Y.}\ \bibnamefont {Ning}}, \bibinfo {author}
  {\bibfnamefont {H.}~\bibnamefont {Su}}, \bibinfo {author} {\bibfnamefont
  {T.}~\bibnamefont {Iitaka}}, \bibinfo {author} {\bibfnamefont
  {T.}~\bibnamefont {Tohyama}}, \ and\ \bibinfo {author} {\bibfnamefont
  {J.-X.}\ \bibnamefont {Zhang}},\ }\href@noop {} {\bibfield  {journal}
  {\bibinfo  {journal} {International Journal of Modern Physics B}\ }\textbf
  {\bibinfo {volume} {31}},\ \bibinfo {pages} {1750217} (\bibinfo {year}
  {2017})}\BibitemShut {NoStop}%
\bibitem [{\citenamefont {{Duan, Xu}}\ \emph {et~al.}(2018)\citenamefont
  {{Duan, Xu}}, \citenamefont {{Wu, Fan}}, \citenamefont {{Chen, Jia}},
  \citenamefont {{Zhang, Peiran}}, \citenamefont {{Liu, Yang}}, \citenamefont
  {{Yuan, Huiqiu}},\ and\ \citenamefont {{Cao, Chao}}}]{DuanCommPhys2018}%
  \BibitemOpen
  \bibfield  {author} {\bibinfo {author} {\bibnamefont {{Duan, Xu}}}, \bibinfo
  {author} {\bibnamefont {{Wu, Fan}}}, \bibinfo {author} {\bibnamefont {{Chen,
  Jia}}}, \bibinfo {author} {\bibnamefont {{Zhang, Peiran}}}, \bibinfo {author}
  {\bibnamefont {{Liu, Yang}}}, \bibinfo {author} {\bibnamefont {{Yuan,
  Huiqiu}}}, \ and\ \bibinfo {author} {\bibnamefont {{Cao, Chao}}},\
  }\href@noop {} {\bibfield  {journal} {\bibinfo  {journal} {Communications
  Physics}\ }\textbf {\bibinfo {volume} {1}},\ \bibinfo {pages} {71} (\bibinfo
  {year} {2018})}\BibitemShut {NoStop}%
\bibitem [{\citenamefont {Huang}\ \emph {et~al.}(2020)\citenamefont {Huang},
  \citenamefont {Lane}, \citenamefont {Cao}, \citenamefont {Zhi}, \citenamefont
  {Liu}, \citenamefont {Matt}, \citenamefont {Kuthanazhi}, \citenamefont
  {Canfield}, \citenamefont {Yarotski}, \citenamefont {Taylor},\ and\
  \citenamefont {Zhu}}]{ZhuPRB2020}%
  \BibitemOpen
  \bibfield  {author} {\bibinfo {author} {\bibfnamefont {Z.}~\bibnamefont
  {Huang}}, \bibinfo {author} {\bibfnamefont {C.}~\bibnamefont {Lane}},
  \bibinfo {author} {\bibfnamefont {C.}~\bibnamefont {Cao}}, \bibinfo {author}
  {\bibfnamefont {G.-X.}\ \bibnamefont {Zhi}}, \bibinfo {author} {\bibfnamefont
  {Y.}~\bibnamefont {Liu}}, \bibinfo {author} {\bibfnamefont {C.~E.}\
  \bibnamefont {Matt}}, \bibinfo {author} {\bibfnamefont {B.}~\bibnamefont
  {Kuthanazhi}}, \bibinfo {author} {\bibfnamefont {P.~C.}\ \bibnamefont
  {Canfield}}, \bibinfo {author} {\bibfnamefont {D.}~\bibnamefont {Yarotski}},
  \bibinfo {author} {\bibfnamefont {A.~J.}\ \bibnamefont {Taylor}}, \ and\
  \bibinfo {author} {\bibfnamefont {J.-X.}\ \bibnamefont {Zhu}},\ }\href
  {\doibase 10.1103/PhysRevB.102.235167} {\bibfield  {journal} {\bibinfo
  {journal} {Phys. Rev. B}\ }\textbf {\bibinfo {volume} {102}},\ \bibinfo
  {pages} {235167} (\bibinfo {year} {2020})}\BibitemShut {NoStop}%
\bibitem [{\citenamefont {Zeng}\ \emph {et~al.}(2015)\citenamefont {Zeng},
  \citenamefont {Fang}, \citenamefont {Chang}, \citenamefont {Chen},
  \citenamefont {Hsieh}, \citenamefont {Bansil}, \citenamefont {Lin},\ and\
  \citenamefont {Fu}}]{zeng2015topological}%
  \BibitemOpen
  \bibfield  {author} {\bibinfo {author} {\bibfnamefont {M.}~\bibnamefont
  {Zeng}}, \bibinfo {author} {\bibfnamefont {C.}~\bibnamefont {Fang}}, \bibinfo
  {author} {\bibfnamefont {G.}~\bibnamefont {Chang}}, \bibinfo {author}
  {\bibfnamefont {Y.-A.}\ \bibnamefont {Chen}}, \bibinfo {author}
  {\bibfnamefont {T.}~\bibnamefont {Hsieh}}, \bibinfo {author} {\bibfnamefont
  {A.}~\bibnamefont {Bansil}}, \bibinfo {author} {\bibfnamefont
  {H.}~\bibnamefont {Lin}}, \ and\ \bibinfo {author} {\bibfnamefont
  {L.}~\bibnamefont {Fu}},\ }\href@noop {} {\bibfield  {journal} {\bibinfo
  {journal} {arXiv preprint arXiv:1504.03492}\ } (\bibinfo {year}
  {2015})}\BibitemShut {NoStop}%
\bibitem [{\citenamefont {Wang}\ \emph {et~al.}(2023)\citenamefont {Wang},
  \citenamefont {Ahn}, \citenamefont {Slager}, \citenamefont {Kushnirenko},
  \citenamefont {Ueland}, \citenamefont {Sapkota}, \citenamefont {Schrunk},
  \citenamefont {Kuthanazhi}, \citenamefont {McQueeney}, \citenamefont
  {Canfield},\ and\ \citenamefont {Kaminski}}]{wang2022multi}%
  \BibitemOpen
  \bibfield  {author} {\bibinfo {author} {\bibfnamefont {L.-L.}\ \bibnamefont
  {Wang}}, \bibinfo {author} {\bibfnamefont {J.}~\bibnamefont {Ahn}}, \bibinfo
  {author} {\bibfnamefont {R.-J.}\ \bibnamefont {Slager}}, \bibinfo {author}
  {\bibfnamefont {Y.}~\bibnamefont {Kushnirenko}}, \bibinfo {author}
  {\bibfnamefont {B.}~\bibnamefont {Ueland}}, \bibinfo {author} {\bibfnamefont
  {A.}~\bibnamefont {Sapkota}}, \bibinfo {author} {\bibfnamefont
  {B.}~\bibnamefont {Schrunk}}, \bibinfo {author} {\bibfnamefont
  {B.}~\bibnamefont {Kuthanazhi}}, \bibinfo {author} {\bibfnamefont
  {R.}~\bibnamefont {McQueeney}}, \bibinfo {author} {\bibfnamefont
  {P.}~\bibnamefont {Canfield}}, \ and\ \bibinfo {author} {\bibfnamefont
  {A.}~\bibnamefont {Kaminski}},\ }\href@noop {} {\bibfield  {journal}
  {\bibinfo  {journal} {Communications Physics}\ } (\bibinfo {year}
  {2023})}\BibitemShut {NoStop}%
\bibitem [{\citenamefont {Zeng}\ \emph {et~al.}(2016)\citenamefont {Zeng},
  \citenamefont {Lou}, \citenamefont {Wu}, \citenamefont {Xu}, \citenamefont
  {Guo}, \citenamefont {Kong}, \citenamefont {Zhong}, \citenamefont {Ma},
  \citenamefont {Fu}, \citenamefont {Richard} \emph
  {et~al.}}]{zeng2016compensated}%
  \BibitemOpen
  \bibfield  {author} {\bibinfo {author} {\bibfnamefont {L.-K.}\ \bibnamefont
  {Zeng}}, \bibinfo {author} {\bibfnamefont {R.}~\bibnamefont {Lou}}, \bibinfo
  {author} {\bibfnamefont {D.-S.}\ \bibnamefont {Wu}}, \bibinfo {author}
  {\bibfnamefont {Q.}~\bibnamefont {Xu}}, \bibinfo {author} {\bibfnamefont
  {P.-J.}\ \bibnamefont {Guo}}, \bibinfo {author} {\bibfnamefont {L.-Y.}\
  \bibnamefont {Kong}}, \bibinfo {author} {\bibfnamefont {Y.-G.}\ \bibnamefont
  {Zhong}}, \bibinfo {author} {\bibfnamefont {J.-Z.}\ \bibnamefont {Ma}},
  \bibinfo {author} {\bibfnamefont {B.-B.}\ \bibnamefont {Fu}}, \bibinfo
  {author} {\bibfnamefont {P.}~\bibnamefont {Richard}},  \emph {et~al.},\
  }\href@noop {} {\bibfield  {journal} {\bibinfo  {journal} {Physical review
  letters}\ }\textbf {\bibinfo {volume} {117}},\ \bibinfo {pages} {127204}
  (\bibinfo {year} {2016})}\BibitemShut {NoStop}%
\bibitem [{\citenamefont {Niu}\ \emph {et~al.}(2016)\citenamefont {Niu},
  \citenamefont {Xu}, \citenamefont {Bai}, \citenamefont {Song}, \citenamefont
  {Shen}, \citenamefont {Xie}, \citenamefont {Sun}, \citenamefont {Huang},
  \citenamefont {Peets},\ and\ \citenamefont {Feng}}]{Niu2016Presence}%
  \BibitemOpen
  \bibfield  {author} {\bibinfo {author} {\bibfnamefont {X.}~\bibnamefont
  {Niu}}, \bibinfo {author} {\bibfnamefont {D.}~\bibnamefont {Xu}}, \bibinfo
  {author} {\bibfnamefont {Y.}~\bibnamefont {Bai}}, \bibinfo {author}
  {\bibfnamefont {Q.}~\bibnamefont {Song}}, \bibinfo {author} {\bibfnamefont
  {X.}~\bibnamefont {Shen}}, \bibinfo {author} {\bibfnamefont {B.}~\bibnamefont
  {Xie}}, \bibinfo {author} {\bibfnamefont {Z.}~\bibnamefont {Sun}}, \bibinfo
  {author} {\bibfnamefont {Y.}~\bibnamefont {Huang}}, \bibinfo {author}
  {\bibfnamefont {D.}~\bibnamefont {Peets}}, \ and\ \bibinfo {author}
  {\bibfnamefont {D.}~\bibnamefont {Feng}},\ }\href@noop {} {\bibfield
  {journal} {\bibinfo  {journal} {Physical Review B}\ }\textbf {\bibinfo
  {volume} {94}},\ \bibinfo {pages} {165163} (\bibinfo {year}
  {2016})}\BibitemShut {NoStop}%
\bibitem [{\citenamefont {Lou}\ \emph {et~al.}(2017)\citenamefont {Lou},
  \citenamefont {Fu}, \citenamefont {Xu}, \citenamefont {Guo}, \citenamefont
  {Kong}, \citenamefont {Zeng}, \citenamefont {Ma}, \citenamefont {Richard},
  \citenamefont {Fang}, \citenamefont {Huang} \emph
  {et~al.}}]{lou2017evidence}%
  \BibitemOpen
  \bibfield  {author} {\bibinfo {author} {\bibfnamefont {R.}~\bibnamefont
  {Lou}}, \bibinfo {author} {\bibfnamefont {B.-B.}\ \bibnamefont {Fu}},
  \bibinfo {author} {\bibfnamefont {Q.}~\bibnamefont {Xu}}, \bibinfo {author}
  {\bibfnamefont {P.-J.}\ \bibnamefont {Guo}}, \bibinfo {author} {\bibfnamefont
  {L.-Y.}\ \bibnamefont {Kong}}, \bibinfo {author} {\bibfnamefont {L.-K.}\
  \bibnamefont {Zeng}}, \bibinfo {author} {\bibfnamefont {J.-Z.}\ \bibnamefont
  {Ma}}, \bibinfo {author} {\bibfnamefont {P.}~\bibnamefont {Richard}},
  \bibinfo {author} {\bibfnamefont {C.}~\bibnamefont {Fang}}, \bibinfo {author}
  {\bibfnamefont {Y.-B.}\ \bibnamefont {Huang}},  \emph {et~al.},\ }\href@noop
  {} {\bibfield  {journal} {\bibinfo  {journal} {Physical Review B}\ }\textbf
  {\bibinfo {volume} {95}},\ \bibinfo {pages} {115140} (\bibinfo {year}
  {2017})}\BibitemShut {NoStop}%
\bibitem [{\citenamefont {Kuroda}\ \emph {et~al.}(2018)\citenamefont {Kuroda},
  \citenamefont {Ochi}, \citenamefont {Suzuki}, \citenamefont {Hirayama},
  \citenamefont {Nakayama}, \citenamefont {Noguchi}, \citenamefont {Bareille},
  \citenamefont {Akebi}, \citenamefont {Kunisada}, \citenamefont {Muro},
  \citenamefont {Watson}, \citenamefont {Kitazawa}, \citenamefont {Haga},
  \citenamefont {Kim}, \citenamefont {Hoesch}, \citenamefont {Shin},
  \citenamefont {Arita},\ and\ \citenamefont {Kondo}}]{Kuroda_2018}%
  \BibitemOpen
  \bibfield  {author} {\bibinfo {author} {\bibfnamefont {K.}~\bibnamefont
  {Kuroda}}, \bibinfo {author} {\bibfnamefont {M.}~\bibnamefont {Ochi}},
  \bibinfo {author} {\bibfnamefont {H.~S.}\ \bibnamefont {Suzuki}}, \bibinfo
  {author} {\bibfnamefont {M.}~\bibnamefont {Hirayama}}, \bibinfo {author}
  {\bibfnamefont {M.}~\bibnamefont {Nakayama}}, \bibinfo {author}
  {\bibfnamefont {R.}~\bibnamefont {Noguchi}}, \bibinfo {author} {\bibfnamefont
  {C.}~\bibnamefont {Bareille}}, \bibinfo {author} {\bibfnamefont
  {S.}~\bibnamefont {Akebi}}, \bibinfo {author} {\bibfnamefont
  {S.}~\bibnamefont {Kunisada}}, \bibinfo {author} {\bibfnamefont
  {T.}~\bibnamefont {Muro}}, \bibinfo {author} {\bibfnamefont {M.~D.}\
  \bibnamefont {Watson}}, \bibinfo {author} {\bibfnamefont {H.}~\bibnamefont
  {Kitazawa}}, \bibinfo {author} {\bibfnamefont {Y.}~\bibnamefont {Haga}},
  \bibinfo {author} {\bibfnamefont {T.~K.}\ \bibnamefont {Kim}}, \bibinfo
  {author} {\bibfnamefont {M.}~\bibnamefont {Hoesch}}, \bibinfo {author}
  {\bibfnamefont {S.}~\bibnamefont {Shin}}, \bibinfo {author} {\bibfnamefont
  {R.}~\bibnamefont {Arita}}, \ and\ \bibinfo {author} {\bibfnamefont
  {T.}~\bibnamefont {Kondo}},\ }\href {\doibase 10.1103/PhysRevLett.120.086402}
  {\bibfield  {journal} {\bibinfo  {journal} {Phys. Rev. Lett.}\ }\textbf
  {\bibinfo {volume} {120}},\ \bibinfo {pages} {086402} (\bibinfo {year}
  {2018})}\BibitemShut {NoStop}%
\bibitem [{\citenamefont {Li}\ \emph {et~al.}(2018)\citenamefont {Li},
  \citenamefont {Wu}, \citenamefont {Wu}, \citenamefont {Cao}, \citenamefont
  {Guo}, \citenamefont {Wu}, \citenamefont {Liu}, \citenamefont {Sun},
  \citenamefont {Cheng}, \citenamefont {Lin} \emph {et~al.}}]{li2018tunable}%
  \BibitemOpen
  \bibfield  {author} {\bibinfo {author} {\bibfnamefont {P.}~\bibnamefont
  {Li}}, \bibinfo {author} {\bibfnamefont {Z.}~\bibnamefont {Wu}}, \bibinfo
  {author} {\bibfnamefont {F.}~\bibnamefont {Wu}}, \bibinfo {author}
  {\bibfnamefont {C.}~\bibnamefont {Cao}}, \bibinfo {author} {\bibfnamefont
  {C.}~\bibnamefont {Guo}}, \bibinfo {author} {\bibfnamefont {Y.}~\bibnamefont
  {Wu}}, \bibinfo {author} {\bibfnamefont {Y.}~\bibnamefont {Liu}}, \bibinfo
  {author} {\bibfnamefont {Z.}~\bibnamefont {Sun}}, \bibinfo {author}
  {\bibfnamefont {C.-M.}\ \bibnamefont {Cheng}}, \bibinfo {author}
  {\bibfnamefont {D.-S.}\ \bibnamefont {Lin}},  \emph {et~al.},\ }\href@noop {}
  {\bibfield  {journal} {\bibinfo  {journal} {Physical Review B}\ }\textbf
  {\bibinfo {volume} {98}},\ \bibinfo {pages} {085103} (\bibinfo {year}
  {2018})}\BibitemShut {NoStop}%
\bibitem [{\citenamefont {Oinuma}\ \emph {et~al.}(2019)\citenamefont {Oinuma},
  \citenamefont {Souma}, \citenamefont {Nakayama}, \citenamefont {Horiba},
  \citenamefont {Kumigashira}, \citenamefont {Yoshida}, \citenamefont {Ochiai},
  \citenamefont {Takahashi},\ and\ \citenamefont {Sato}}]{SatoCeBi}%
  \BibitemOpen
  \bibfield  {author} {\bibinfo {author} {\bibfnamefont {H.}~\bibnamefont
  {Oinuma}}, \bibinfo {author} {\bibfnamefont {S.}~\bibnamefont {Souma}},
  \bibinfo {author} {\bibfnamefont {K.}~\bibnamefont {Nakayama}}, \bibinfo
  {author} {\bibfnamefont {K.}~\bibnamefont {Horiba}}, \bibinfo {author}
  {\bibfnamefont {H.}~\bibnamefont {Kumigashira}}, \bibinfo {author}
  {\bibfnamefont {M.}~\bibnamefont {Yoshida}}, \bibinfo {author} {\bibfnamefont
  {A.}~\bibnamefont {Ochiai}}, \bibinfo {author} {\bibfnamefont
  {T.}~\bibnamefont {Takahashi}}, \ and\ \bibinfo {author} {\bibfnamefont
  {T.}~\bibnamefont {Sato}},\ }\href {\doibase 10.1103/PhysRevB.100.125122}
  {\bibfield  {journal} {\bibinfo  {journal} {Phys. Rev. B}\ }\textbf {\bibinfo
  {volume} {100}},\ \bibinfo {pages} {125122} (\bibinfo {year}
  {2019})}\BibitemShut {NoStop}%
\bibitem [{\citenamefont {Sakhya}\ \emph
  {et~al.}(2022{\natexlab{a}})\citenamefont {Sakhya}, \citenamefont {Kumar},
  \citenamefont {Pramanik}, \citenamefont {Pandeya}, \citenamefont {Verma},
  \citenamefont {Singh}, \citenamefont {Datta}, \citenamefont {Sasmal},
  \citenamefont {Mondal}, \citenamefont {Schwier} \emph
  {et~al.}}]{sakhya2022behavior}%
  \BibitemOpen
  \bibfield  {author} {\bibinfo {author} {\bibfnamefont {A.~P.}\ \bibnamefont
  {Sakhya}}, \bibinfo {author} {\bibfnamefont {S.}~\bibnamefont {Kumar}},
  \bibinfo {author} {\bibfnamefont {A.}~\bibnamefont {Pramanik}}, \bibinfo
  {author} {\bibfnamefont {R.~P.}\ \bibnamefont {Pandeya}}, \bibinfo {author}
  {\bibfnamefont {R.}~\bibnamefont {Verma}}, \bibinfo {author} {\bibfnamefont
  {B.}~\bibnamefont {Singh}}, \bibinfo {author} {\bibfnamefont
  {S.}~\bibnamefont {Datta}}, \bibinfo {author} {\bibfnamefont
  {S.}~\bibnamefont {Sasmal}}, \bibinfo {author} {\bibfnamefont
  {R.}~\bibnamefont {Mondal}}, \bibinfo {author} {\bibfnamefont {E.~F.}\
  \bibnamefont {Schwier}},  \emph {et~al.},\ }\href@noop {} {\bibfield
  {journal} {\bibinfo  {journal} {Physical Review B}\ }\textbf {\bibinfo
  {volume} {106}},\ \bibinfo {pages} {085132} (\bibinfo {year}
  {2022}{\natexlab{a}})}\BibitemShut {NoStop}%
\bibitem [{\citenamefont {Sakhya}\ \emph
  {et~al.}(2022{\natexlab{b}})\citenamefont {Sakhya}, \citenamefont {Wang},
  \citenamefont {Kabir}, \citenamefont {Huang}, \citenamefont {Hosen},
  \citenamefont {Singh}, \citenamefont {Regmi}, \citenamefont {Dhakal},
  \citenamefont {Dimitri}, \citenamefont {Sprague}, \citenamefont {Smith},
  \citenamefont {Bauer}, \citenamefont {Ronning}, \citenamefont {Bansil},\ and\
  \citenamefont {Neupane}}]{SakhyaNdSb2022}%
  \BibitemOpen
  \bibfield  {author} {\bibinfo {author} {\bibfnamefont {A.~P.}\ \bibnamefont
  {Sakhya}}, \bibinfo {author} {\bibfnamefont {B.}~\bibnamefont {Wang}},
  \bibinfo {author} {\bibfnamefont {F.}~\bibnamefont {Kabir}}, \bibinfo
  {author} {\bibfnamefont {C.-Y.}\ \bibnamefont {Huang}}, \bibinfo {author}
  {\bibfnamefont {M.~M.}\ \bibnamefont {Hosen}}, \bibinfo {author}
  {\bibfnamefont {B.}~\bibnamefont {Singh}}, \bibinfo {author} {\bibfnamefont
  {S.}~\bibnamefont {Regmi}}, \bibinfo {author} {\bibfnamefont
  {G.}~\bibnamefont {Dhakal}}, \bibinfo {author} {\bibfnamefont
  {K.}~\bibnamefont {Dimitri}}, \bibinfo {author} {\bibfnamefont
  {M.}~\bibnamefont {Sprague}}, \bibinfo {author} {\bibfnamefont
  {R.}~\bibnamefont {Smith}}, \bibinfo {author} {\bibfnamefont {E.~D.}\
  \bibnamefont {Bauer}}, \bibinfo {author} {\bibfnamefont {F.}~\bibnamefont
  {Ronning}}, \bibinfo {author} {\bibfnamefont {A.}~\bibnamefont {Bansil}}, \
  and\ \bibinfo {author} {\bibfnamefont {M.}~\bibnamefont {Neupane}},\ }\href
  {\doibase 10.1103/PhysRevB.106.235119} {\bibfield  {journal} {\bibinfo
  {journal} {Phys. Rev. B}\ }\textbf {\bibinfo {volume} {106}},\ \bibinfo
  {pages} {235119} (\bibinfo {year} {2022}{\natexlab{b}})}\BibitemShut
  {NoStop}%
\bibitem [{\citenamefont {Li}\ \emph {et~al.}(1996)\citenamefont {Li},
  \citenamefont {Haga}, \citenamefont {Shida}, \citenamefont {Suzuki},\ and\
  \citenamefont {Kwon}}]{li1996electrical}%
  \BibitemOpen
  \bibfield  {author} {\bibinfo {author} {\bibfnamefont {D.}~\bibnamefont
  {Li}}, \bibinfo {author} {\bibfnamefont {Y.}~\bibnamefont {Haga}}, \bibinfo
  {author} {\bibfnamefont {H.}~\bibnamefont {Shida}}, \bibinfo {author}
  {\bibfnamefont {T.}~\bibnamefont {Suzuki}}, \ and\ \bibinfo {author}
  {\bibfnamefont {Y.}~\bibnamefont {Kwon}},\ }\href@noop {} {\bibfield
  {journal} {\bibinfo  {journal} {Physical Review B}\ }\textbf {\bibinfo
  {volume} {54}},\ \bibinfo {pages} {10483} (\bibinfo {year}
  {1996})}\BibitemShut {NoStop}%
\bibitem [{\citenamefont {Dwari}\ \emph {et~al.}(2023)\citenamefont {Dwari},
  \citenamefont {Sasmal}, \citenamefont {Dan}, \citenamefont {Maity},
  \citenamefont {Saini}, \citenamefont {Kulkarni}, \citenamefont {Banik},
  \citenamefont {Verma}, \citenamefont {Singh},\ and\ \citenamefont
  {Thamizhavel}}]{dwari2023large}%
  \BibitemOpen
  \bibfield  {author} {\bibinfo {author} {\bibfnamefont {G.}~\bibnamefont
  {Dwari}}, \bibinfo {author} {\bibfnamefont {S.}~\bibnamefont {Sasmal}},
  \bibinfo {author} {\bibfnamefont {S.}~\bibnamefont {Dan}}, \bibinfo {author}
  {\bibfnamefont {B.}~\bibnamefont {Maity}}, \bibinfo {author} {\bibfnamefont
  {V.}~\bibnamefont {Saini}}, \bibinfo {author} {\bibfnamefont
  {R.}~\bibnamefont {Kulkarni}}, \bibinfo {author} {\bibfnamefont
  {S.}~\bibnamefont {Banik}}, \bibinfo {author} {\bibfnamefont
  {R.}~\bibnamefont {Verma}}, \bibinfo {author} {\bibfnamefont
  {B.}~\bibnamefont {Singh}}, \ and\ \bibinfo {author} {\bibfnamefont
  {A.}~\bibnamefont {Thamizhavel}},\ }\href@noop {} {\bibfield  {journal}
  {\bibinfo  {journal} {Physical Review B}\ }\textbf {\bibinfo {volume}
  {107}},\ \bibinfo {pages} {235117} (\bibinfo {year} {2023})}\BibitemShut
  {NoStop}%
\bibitem [{\citenamefont {Hulliger}(1980)}]{hulliger1980magnetic}%
  \BibitemOpen
  \bibfield  {author} {\bibinfo {author} {\bibfnamefont {F.}~\bibnamefont
  {Hulliger}},\ }\href@noop {} {\bibfield  {journal} {\bibinfo  {journal}
  {Journal of Magnetism and Magnetic Materials}\ }\textbf {\bibinfo {volume}
  {15}},\ \bibinfo {pages} {1243} (\bibinfo {year} {1980})}\BibitemShut
  {NoStop}%
\bibitem [{\citenamefont {Zhao}\ \emph {et~al.}(2022)\citenamefont {Zhao},
  \citenamefont {Rahman}, \citenamefont {Liu}, \citenamefont {Meng},
  \citenamefont {Ling}, \citenamefont {Han}, \citenamefont {Xi}, \citenamefont
  {Tong}, \citenamefont {Xu}, \citenamefont {Tian} \emph
  {et~al.}}]{zhao2022multiple}%
  \BibitemOpen
  \bibfield  {author} {\bibinfo {author} {\bibfnamefont {J.}~\bibnamefont
  {Zhao}}, \bibinfo {author} {\bibfnamefont {A.}~\bibnamefont {Rahman}},
  \bibinfo {author} {\bibfnamefont {W.}~\bibnamefont {Liu}}, \bibinfo {author}
  {\bibfnamefont {F.}~\bibnamefont {Meng}}, \bibinfo {author} {\bibfnamefont
  {L.}~\bibnamefont {Ling}}, \bibinfo {author} {\bibfnamefont {Y.}~\bibnamefont
  {Han}}, \bibinfo {author} {\bibfnamefont {C.}~\bibnamefont {Xi}}, \bibinfo
  {author} {\bibfnamefont {W.}~\bibnamefont {Tong}}, \bibinfo {author}
  {\bibfnamefont {L.}~\bibnamefont {Xu}}, \bibinfo {author} {\bibfnamefont
  {Z.}~\bibnamefont {Tian}},  \emph {et~al.},\ }\href@noop {} {\bibfield
  {journal} {\bibinfo  {journal} {Physical Review B}\ }\textbf {\bibinfo
  {volume} {106}},\ \bibinfo {pages} {224412} (\bibinfo {year}
  {2022})}\BibitemShut {NoStop}%
\bibitem [{\citenamefont {Nereson}\ and\ \citenamefont
  {Struebing}(1972)}]{nereson1972neutron}%
  \BibitemOpen
  \bibfield  {author} {\bibinfo {author} {\bibfnamefont {N.}~\bibnamefont
  {Nereson}}\ and\ \bibinfo {author} {\bibfnamefont {V.}~\bibnamefont
  {Struebing}},\ }in\ \href@noop {} {\emph {\bibinfo {booktitle} {AIP
  Conference Proceedings}}},\ Vol.~\bibinfo {volume} {5}\ (\bibinfo
  {organization} {American Institute of Physics},\ \bibinfo {year} {1972})\
  pp.\ \bibinfo {pages} {1385--1389}\BibitemShut {NoStop}%
\bibitem [{\citenamefont {Li}(1955)}]{li1955magnetic}%
  \BibitemOpen
  \bibfield  {author} {\bibinfo {author} {\bibfnamefont {Y.-Y.}\ \bibnamefont
  {Li}},\ }\href@noop {} {\bibfield  {journal} {\bibinfo  {journal} {Physical
  Review}\ }\textbf {\bibinfo {volume} {100}},\ \bibinfo {pages} {627}
  (\bibinfo {year} {1955})}\BibitemShut {NoStop}%
\bibitem [{\citenamefont {Canfield}\ and\ \citenamefont
  {Fisk}(1992)}]{Canfield1992Growth}%
  \BibitemOpen
  \bibfield  {author} {\bibinfo {author} {\bibfnamefont {P.~C.}\ \bibnamefont
  {Canfield}}\ and\ \bibinfo {author} {\bibfnamefont {Z.}~\bibnamefont
  {Fisk}},\ }\href {\doibase 10.1080/13642819208215073} {\bibfield  {journal}
  {\bibinfo  {journal} {Philosophical Magazine Part B}\ }\textbf {\bibinfo
  {volume} {65}},\ \bibinfo {pages} {1117} (\bibinfo {year}
  {1992})}\BibitemShut {NoStop}%
\bibitem [{\citenamefont {Canfield}\ \emph {et~al.}(2016)\citenamefont
  {Canfield}, \citenamefont {Kong}, \citenamefont {Kaluarachchi},\ and\
  \citenamefont {Jo}}]{Canfield2016Use}%
  \BibitemOpen
  \bibfield  {author} {\bibinfo {author} {\bibfnamefont {P.~C.}\ \bibnamefont
  {Canfield}}, \bibinfo {author} {\bibfnamefont {T.}~\bibnamefont {Kong}},
  \bibinfo {author} {\bibfnamefont {U.~S.}\ \bibnamefont {Kaluarachchi}}, \
  and\ \bibinfo {author} {\bibfnamefont {N.~H.}\ \bibnamefont {Jo}},\
  }\href@noop {} {\bibfield  {journal} {\bibinfo  {journal} {Philosophical
  Magazine}\ }\textbf {\bibinfo {volume} {96}},\ \bibinfo {pages} {84}
  (\bibinfo {year} {2016})}\BibitemShut {NoStop}%
\bibitem [{\citenamefont {URL:}()}]{FrittURL}%
  \BibitemOpen
  \bibfield  {author} {\bibinfo {author} {\bibnamefont {URL:}},\ }\href@noop {}
  {\bibinfo  {journal} {https://lspceramics.com/canfield-crucible-sets-2/}\
  }\BibitemShut {NoStop}%
\bibitem [{\citenamefont {Canfield}(2019)}]{Canfield_2019}%
  \BibitemOpen
\bibfield  {journal} {  }\bibfield  {author} {\bibinfo {author} {\bibfnamefont
  {P.~C.}\ \bibnamefont {Canfield}},\ }\href {\doibase
  10.1088/1361-6633/ab514b} {\bibfield  {journal} {\bibinfo  {journal} {Reports
  on Progress in Physics}\ }\textbf {\bibinfo {volume} {83}},\ \bibinfo {pages}
  {016501} (\bibinfo {year} {2019})}\BibitemShut {NoStop}%
\bibitem [{\citenamefont {Jiang}\ \emph {et~al.}(2014)\citenamefont {Jiang},
  \citenamefont {Mou}, \citenamefont {Wu}, \citenamefont {Huang}, \citenamefont
  {McMillen}, \citenamefont {Kolis}, \citenamefont {Giesber}, \citenamefont
  {Egan},\ and\ \citenamefont {Kaminski}}]{jiang2014tunable}%
  \BibitemOpen
  \bibfield  {author} {\bibinfo {author} {\bibfnamefont {R.}~\bibnamefont
  {Jiang}}, \bibinfo {author} {\bibfnamefont {D.}~\bibnamefont {Mou}}, \bibinfo
  {author} {\bibfnamefont {Y.}~\bibnamefont {Wu}}, \bibinfo {author}
  {\bibfnamefont {L.}~\bibnamefont {Huang}}, \bibinfo {author} {\bibfnamefont
  {C.~D.}\ \bibnamefont {McMillen}}, \bibinfo {author} {\bibfnamefont
  {J.}~\bibnamefont {Kolis}}, \bibinfo {author} {\bibfnamefont {H.~G.}\
  \bibnamefont {Giesber}}, \bibinfo {author} {\bibfnamefont {J.~J.}\
  \bibnamefont {Egan}}, \ and\ \bibinfo {author} {\bibfnamefont
  {A.}~\bibnamefont {Kaminski}},\ }\href {\doibase
  http://dx.doi.org/10.1063/1.4867517} {\bibfield  {journal} {\bibinfo
  {journal} {Review of Scientific Instruments}\ }\textbf {\bibinfo {volume}
  {85}},\ \bibinfo {eid} {033902} (\bibinfo {year} {2014})}\BibitemShut
  {NoStop}%
\end{thebibliography}%

\end{document}